\documentclass[useAMS,usenatbib]{mn2e}
\usepackage{graphicx}
\usepackage{amssymb}
\usepackage{color}
\usepackage{tabularx}
\usepackage{multirow}
\usepackage{bm}
\usepackage{amssymb}
\usepackage{macros}
\usepackage[usenames,dvipsnames,svgnames,table]{xcolor}
\usepackage{hyperref}
\definecolor{darkblue}{rgb}{0.0,0.0,0.5}
\definecolor{darkred}{rgb}{0.5,0.0,0.0}
\hypersetup{colorlinks,breaklinks,
            linkcolor=red,urlcolor=darkblue,
            anchorcolor=darkblue,citecolor=darkblue}
\usepackage{graphicx}
\usepackage{amsmath}
\usepackage{epstopdf}

\def\ba{\begin{eqnarray}}
\def\ea{\end{eqnarray}}
\def\be{\begin{equation}}
\def\ee{\end{equation}}
\usepackage{amsmath}

\newcommand{\appropto}{\mathrel{\vcenter{
  \offinterlineskip\halign{\hfil$##$\cr
    \propto\cr\noalign{\kern2pt}\sim\cr\noalign{\kern-2pt}}}}}

\title[On the statistics of interpulse radio pulsars]
{Statistics of interpulse radio pulsars -- 
the key to solving the alignment/counter-alignment problem}
\author[Arzamasskiy, Beskin \& Pirov]{L.I. Arzamasskiy$^1$\thanks{E-mail: leva@astro.princeton.edu, beskin@lpi.ru}, 
V.S. Beskin$^{2,3\star}$\footnote[0]~ and K.K. Pirov$^{3}$ \\
$^{1}$Department of Astrophysical Sciences, Peyton Hall, Princeton University, Princeton, NJ 08544, USA \\
$^{2}$P.N.Lebedev Physical Institute, Leninsky prosp., 53, Moscow, 119991, Russia\\
$^{3}$Moscow Institute of Physics and Technology, Dolgoprudny, Institutsky per., 9, 
Moscow region, 141700, Russia\\
}

\begin{document}

\date{Accepted. Received; in original form}

\pagerange{\pageref{firstpage}--\pageref{lastpage}} \pubyear{2016}

\maketitle

\label{firstpage}

\begin{abstract}
At present, there are theoretical models of radio pulsar evolution which predict both the 
alignment, i.e., evolution of inclination angle $\chi$ between magnetic and rotational 
axes to $0^{\circ}$, and its counter-alignment, i.e., evolution to $90^{\circ}$. At the same time, both 
models well describe the pulsar distribution on $P$--$\dot P$ diagram. For this reason, up 
to now it was impossible to determine the braking mechanisms since it was rather difficult to 
estimate inclination angle evolution on the basis of observation. In this paper we demonstrate that 
statistics of interpulse pulsars can give us the key to solve alignment/counter-alignment problem 
as the number of interpulse pulsars (both, having $\chi \sim 0^{\circ}$ and $\chi \sim 90^{\circ}$) 
drastically depends on evolution of inclination angle.

\end{abstract}



\section{Introduction}
\label{sect:intro}

Almost fifty years after radio pulsars discovery the problem of neutron star energy loss 
still remains unsolved~\citep{1977puls.book.....M, 1977puls.book.....S}. In particular, 
evolution of the inclination angle $\chi$ between magnetic and rotational axes is still 
unknown. At present, there are theoretical models which predict both inclination angle 
evolution to $0^{\circ}$, i.e., alignment~\citep{1970ApJ...159L..81D, 1970ApJ...160L..11G, 1985ApJ...299..706G, 2014MNRAS.441.1879P} and its 
evolution to $90^{\circ}$, i.e., counter-alignment~\citep{BGI}. Both models are good in describing $P$--$\dot P$  diagram (which is directly observed), but give completely different answers to the question of inclination angle evolution (for which we have very little observations). 

There were many attempts to resolve the issue by analyzing statistical distribution 
of radio pulsars~\citep{1990ApJ...352..247R, 1998MNRAS.298..625T,2006ApJ...643..332F, 2008MNRAS.387.1755W, 2010MNRAS.402.1317Y, 2014MNRAS.443.1891G}. In particular, it was found 
both directly (i.e., by the analysis of the $\chi$ distribution) and indirectly (i.e., 
from the analysis of the observed pulse width) that statistically the inclination angle 
$\chi$ decreases with period $P$ as the dynamical age $\tau_{\rm D} = P/{\dot P}$ increases. 
At first glance, these results definitely speak in favor of alignment mechanism. However, as 
was demonstrated by~\citet{BGI}, the average inclination angle of pulsar population, 
$\langle\chi\rangle (\tau_{\rm D})$, computed for observed pulsars can 
\emph{decrease} even if inclination angles of individual pulsars \emph{increases} with time.

Indeed, for given values of pulsar period $P$ and  magnetic field $B$ the secondary pair 
production over magnetic polar cap is suppressed at angles $\chi$ close to $90^{\circ}$, when 
magnetic dipole is nearly orthogonal to the rotational axis. This is because the
Goldreich-Julian charge density $\rho_{\rm GJ} \approx \Omega B \cos \chi/(2 \pi c)$  is 
significantly reduces at such angles. This in turn leads to decrease in electric 
potential drop near the surface of neutron star and suppression of 
the secondary particles production. Because of relation between pulsar 
extinction line and $\chi$, the average inclination angles of observed populations 
can decrease along dynamical age increase. Detailed analysis, already carried out 
by ~\citet{1984Ap&SS.102..301B} and \citet{2005AstL...31..263B}, on the basis of a kinetic equation
describing the distribution of pulsars provided quantitative proof this picture.

Recently, by analyzing 45 years of observational data for 
the Crab pulsar, \citet{2013Sci...342..598L} found that the separation between the main pulse and interpulse increases at the rate of $~0.6^\circ$ per century (implying similar growth of $\chi$). Even though it argues in favor of counter-alignment model, as it was recently shown by \citet{2015MNRAS.453.3540A} and \citet{2015MNRAS.451..695Z}, the data can be explained with alignment model as well, if precession with characteristic timescale of $\sim 100$ years is considered.

\begin{figure*}
\centering
\includegraphics[scale=0.7]{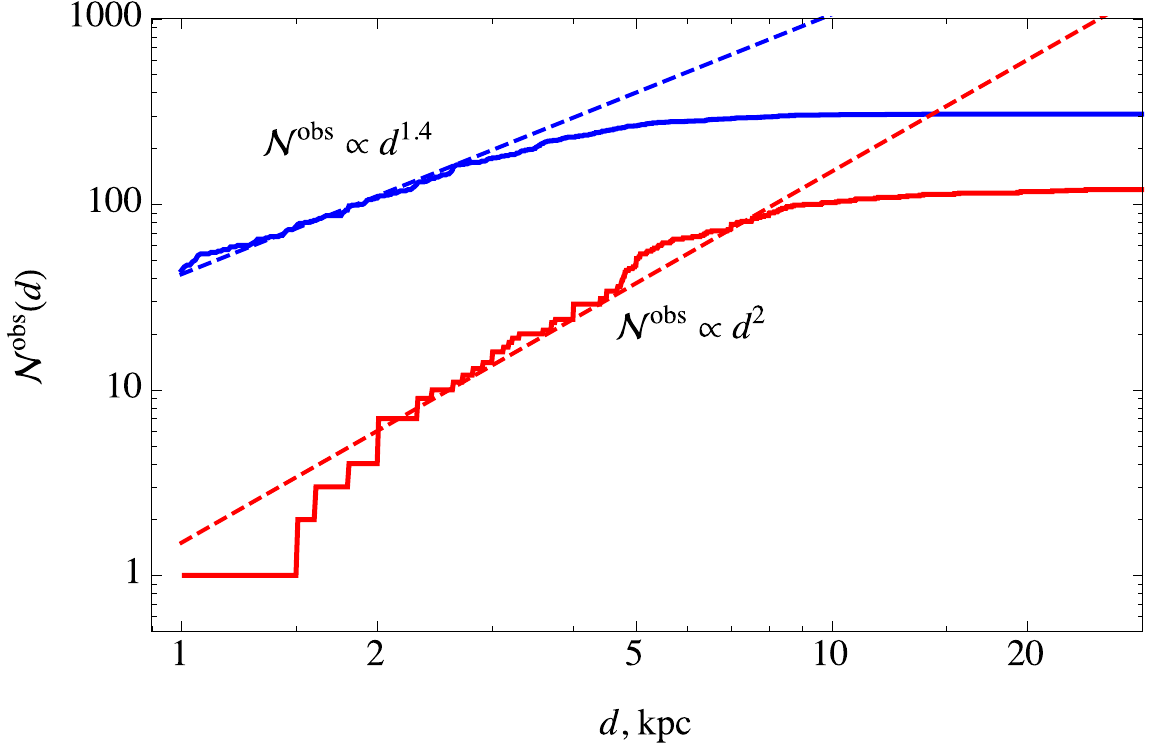}~~~~~\includegraphics[scale=0.7]{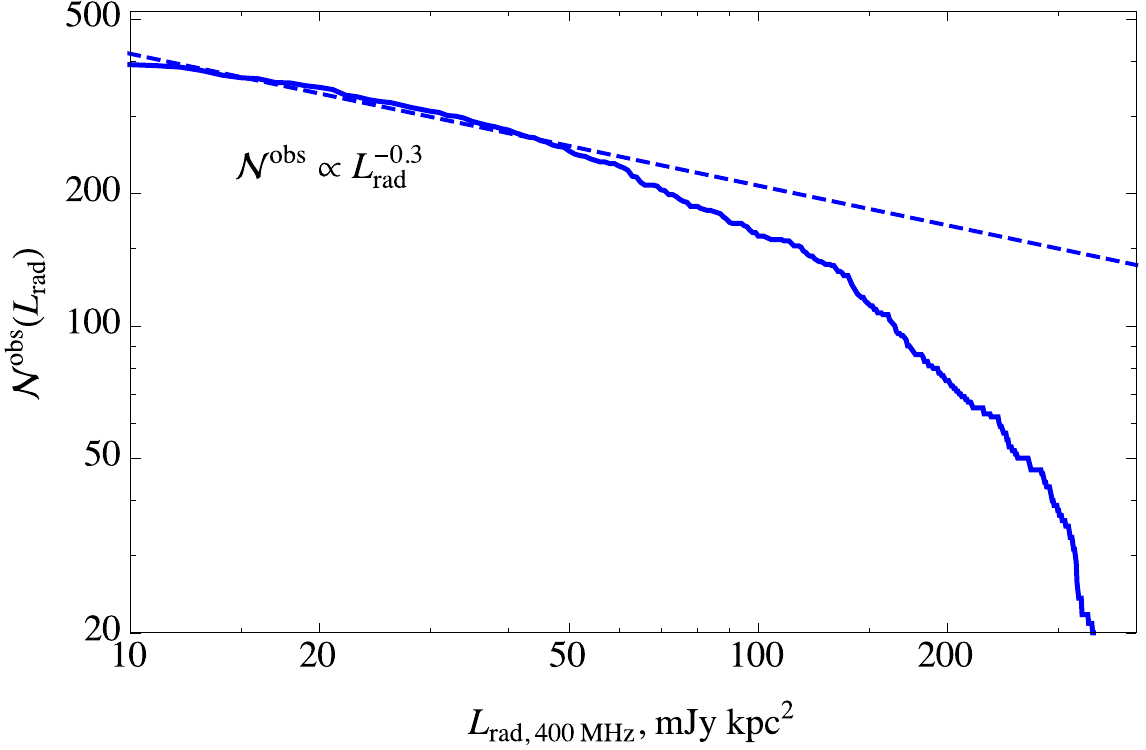}
\caption{
\textbf{[Left panel]:} Spatial distribution of radio pulsars ${\cal N}^{\rm obs}(d^{\rm obs}<d)$.
Upper (blue) curve corresponds to the
the main population of pulsars with luminosity $L_{\rm rad} < 400$ mJy kpc$^2$ measured in 400 MHz waveband; lower (red) curve corresponds to the brightest pulsars
with radio luminosity is $L_{\rm rad} > 400$ mJy kpc$^2$. Brightest pulsars have distribution function ${\cal N}^{\rm obs} \propto d^2$ consistent with homogeneously distributed pulsars in the galactic disk. However, the main population of pulsars has different power-law ${\cal N}^{\rm obs} \propto d^{1.4}$, which we explain by considering luminosity distribution function on the right panel. \textbf{[Right panel]:} Luminosity distribution function in 400 MHz waveband ${\cal N}^{\rm obs} (L^{\rm obs}>L_{\rm rad})$. At small luminosities it has an approximate power-law behavior ${\cal N}^{\rm obs} \propto L_{\rm rad}^{-0.3}$, which allows to explain the behavior of pulsar distribution over distances via eqn. (\ref{Dobs1}).
} 
\label{fig03}
\end{figure*}

Thus, one can conclude that at present there is no common point of view on the evolution of 
inclination angle $\chi$ of radio pulsars. On the other hand, it is quite clear, 
that inclination angle $\chi$ is a key hidden parameter and without taking it into account, 
it is impossible to develop consistent theory  of radio pulsar evolution. 

The aim of this paper is to resolve alignment/counter-alignment problem by analyzing statistical properties of interpulse pulsars, as the number of such pulsars (both for 
$\chi \sim 0^{\circ}$ and $\chi \sim 90^{\circ}$) mainly depends upon the evolution of 
inclination angle. 

The paper is organized as follows. Sect. \ref{sect:obs} is devoted to the analysis of the observational 
data which gives us necessary information about the birth distribution of radio pulsars as well as the visibility function. Here we also 
give the full list of interpulse pulsars. In Sect. \ref{sect:torques} we discuss two main evolution 
theories predicting alignment and counter-alignment evolution. In Sect. \ref{sect:solution} we describe the 
details of our population synthesis based on kinetic equation approach. Finally, in 
Sect. \ref{sect:results} the main results of our consideration are formulated.

\section{Relevant Observations}
\label{sect:obs}

In this section we gather observational constrains on pulsar distribution function. Throughout the paper we use the following notation. We refer to the real distribution function (e.g., the distribution function of \emph{all} pulsars including ones which are not observed) over parameter $f$ as $N(f)$. The observed distribution function is different from the real one due to several selection effects. We refer to such function as $N^{\rm obs}(f)$. When describing observations, we often make use of the integrated observed distribution function $\mathcal{N}^{\rm obs}(f) \equiv \int^f N^{\rm obs}(f'){\rm d}f'$.

\subsection{Spacial distribution}
\label{Obsd}

To start with, we need to make some preliminary remarks concerning general properties 
of the radio pulsar statistical distribution. It helps us determine both the visibility 
function $V^{\rm vis}(P,\chi)$ as well as the birth distribution $Q(P,\chi)$ of radio pulsars. In this subsection we analyze spatial distribution of radio pulsars in the galactic disk.

Fig.~\ref{fig03} (left panel) shows observed spatial distribution function of radio pulsars. We divide all pulsars into two groups: the main population (blue, upper curve) which have radio luminosities $L_{\rm rad} < 400$ mJy kpc$^2$ in 400 MHz waveband, and the brightest ones which have $L_{\rm rad} > 400$ mJy kpc$^2$ (red, lower curve){\footnote{Here we use 
\url{http://www.atnf.csiro.au/people/pulsar/psrcat/} ATNF pulsar catalogue~\citep{2005AJ....129.1993M}}.}. Only pulsars with known $d$ and $L_{\rm rad}$ are taken into account.

As one can see, brightest sources shows reasonable integral distribution 
${\cal N}^{\rm obs}(d) = 2 \pi \int_{0}^{d} N(d') d'\, {\rm d}d'$:
\begin{equation}
{\cal N}^{\rm obs}_{\rm bright}(d) \propto d^{\,2.0}
\label{Ddbright}
\end{equation}
in line with homogeneous distribution of neutron stars within the galactic disk. On the other 
hand, main population demonstrates conspicuous deviation 
\begin{equation}
{\cal N}^{\rm obs}_{\rm main}(d) \propto d^{\, 1.4}.
\label{Dd}
\end{equation}

This disagreement can be easily explained if we include into consideration the 
luminosity visibility function $V^{\rm vis}_{\rm lum}$ implying that the 
receiver with sensitivity $S$ cannot detect distant radio sources with 
$L_{\rm rad} < 4 \pi S d^2$. Indeed, as shown on Fig.~\ref{fig03} (right panel), visible integral
radio luminosity distribution of radio pulsars with $L_{\rm rad} < 400$ mJy kpc$^2$ 
has a power-law dependence at small luminosities
\begin{equation}
{\cal N}^{\rm obs}(L_{\rm rad}) \propto L_{\rm rad}^{-0.3},
\end{equation}
corresponding to differential distribution
\begin{equation}
N^{\rm obs}(L_{\rm rad}) \propto L_{\rm rad}^{-1.3}.
\label{DLum}
\end{equation}
Providing a theoretical prediction for the spatial distribution function
\begin{equation}
{\cal N}_{\rm th}^{\rm obs}(d) = 2 \pi \int_0^d l {\rm d}l \int_{4 \pi S l^2}^{\infty} N^{\rm obs}(L_{\rm rad}) \,
{\rm d}L_{\rm rad} \propto d^{1.4},
\label{Dobs1}
\end{equation}
we obtain a nice agreement with the observed distribution~(\ref{Dd}). 

Thus, one can conclude that the visible spacial distribution of radio pulsars is compatible with 
their homogeneous distribution within the Galactic disk. For this reason, below we do not include into
consideration possible correlations connecting pulsar velocities, their $z$-distribution in the Galactic disk, etc.

\subsection{Angular distribution}
\label{ObsX}

Further, let us try to evaluate the dependence of the distribution of radio pulsars on the inclination angle $\chi$. Unfortunately, as on today, the determination of  angle $\chi$ by 
analyzing the swing of the linear polarization position angle~\citep{1998MNRAS.298..625T, 2011MNRAS.414.1314M, 2013ARep...57..833M} has some uncertainties, so different authors give different values of inclination angle. Moreover, the number of pulsars with well-determined inclination angles $\chi$
is still rather low (approx. 100--200), thus preventing us to discuss in detail 
their statistical properties. 

For this reason herein we use approach proposed by~\citet{1990ApJ...352..247R} and~\citet{2012MNRAS.424.1762M} which 
allows us to evaluate the inclination angle for individual pulsar from its 
observed width of mean profile $W^{\rm obs}_{r}$. Indeed, if $W_{0}$ is an 
intrinsic width of directivity pattern, then observed width for $\chi > W_{0}$ 
will be equal to
\begin{equation}
W^{\rm obs}_{r} = \frac{W_{0}}{\sin\chi}.
\label{Wobs1}
\end{equation}
As it was found by~\citet{1990ApJ...352..247R, 1993ApJ...405..285R}, and~\citet{2012MNRAS.424.1762M}, one has different values of $W_0$ for conal and core components of emission. In this paper, we mainly use the value of $W_0$ corresponding to conal component (as was also used in \citealt{2008MNRAS.387.1755W}):
\begin{equation}
W_{0} = \frac{5.4^\circ}{\sqrt{P}}.
\label{Wobs2}
\end{equation}
Here factor $P^{-1/2}$ (where $P$ is in seconds) corresponds to the clear period dependence upon 
open magnetic field lines which just determines the diagram width. As a result, relations 
(\ref{Wobs1})--(\ref{Wobs2}) allow us to evaluate angular distribution of radio pulsars on 
much richer statistics.

\begin{figure}
\centering
\includegraphics[scale=0.7]{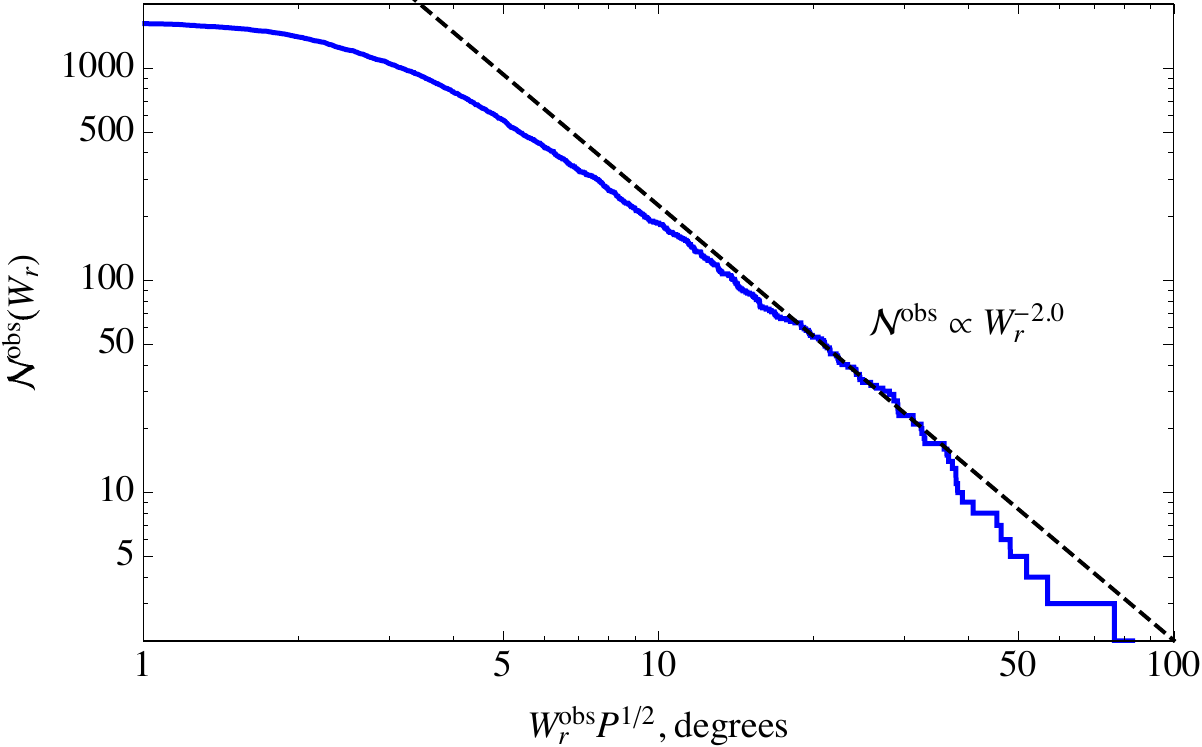}
\caption{
Observed integrated window width distribution 
${\cal N}^{\rm obs}(W_{r}) \propto (W_{r})^{-2.0 \pm 0.2}$ 
determined from statistics of the mean profile width $W_r = W_r^{\rm obs} P^{1/2}$ 
(taken from ATNF pulsar catalogue at 50\% intensity level). One can see, that it has approximate power-law dependence for small inclination angles (large window widths) with index -2, implying $N^{\rm obs} (\chi) \propto \chi$ at small  angles $\chi$. 
} 
\label{fig06}
\end{figure}

As is shown on Fig.~\ref{fig06}, the observed window width distribution 
${\cal N}^{\rm obs} =\int N(W_{r}) {\rm d}W_{r}$  for 
$W_{r} = W_r^{\rm obs}P^{1/2} < 35^{\circ}$ features a power-law dependence 
corresponding to differential distribution
\begin{equation}
N^{\rm obs}(W_{r}) \propto (W_{r})^{-3.0}.
\label{Wobs3}
\end{equation}
 As
\begin{equation}
N^{\rm obs}(\chi) 
= N^{\rm obs}(W_{r})\frac{{\rm d}W_{r}}{{\rm d}\chi},
\label{Wobs4}
\end{equation}
one can conclude that observed angular distribution $N^{\rm obs}(\chi)$ at small angles 
shall be proportional to $\chi$ 
\begin{equation}
N^{\rm obs}(\chi) \propto \chi,
\label{Wobs5}
\end{equation}
which is in a good agreement with observations~\citep{1998MNRAS.298..625T, 2012MNRAS.424.1762M}.

For  $\chi > W_{0}$ the beaming visibility function $V^{\rm vis}_{\rm beam}$
(which takes into account that the observer must be located within the directivity pattern of the radio beam) 
can be written as $V^{\rm vis}_{\rm beam} =  \sin\chi W^{\rm obs}_{r}$.
Accordingly, if one can put $N^{\rm obs}(\chi)=  V^{\rm vis}_{\rm beam}(\chi) N(\chi)$, the real 
distribution function $N(\chi)$ for small angles $\chi$ shall be approximately constant: 
\begin{equation}
N(\chi) \approx {\rm const} \quad {\rm (small~angles)}.
\label{const}
\end{equation} 

Thus, we come to conclusion that when analyzing observed distribution of radio pulsars, 
it is necessary to involve the beaming visibility function $V^{\rm vis}_{\rm beam}$ which 
in general case can be approximately formulated as (see more accurate definition in Sect.~\ref{sect:vis_f}): 
\begin{eqnarray} 
V^{\rm vis}_{\rm beam} = 
\begin{cases} 
 \sin\chi W_{0}, &   \chi > W_{0}, \\ 
 W^{2}_{0}, &   \chi < W_{0}.
\end{cases}
\label{Vdiag}
\end{eqnarray}
It is interesting that the break for $W_{r} > 35^{\circ}$ (see Fig.~\ref{fig06})
just corresponds to inclination angles $\chi < W_{0}$ when the lower expression
in (\ref{Vdiag}) is to be used. 

\subsection{Visibility function}
\label{sect:vis_f}

The observed distribution function $N^{\rm obs}$ of radio pulsars deviates from the real distribution function $N$. That difference comes from two main effects.

The first one comes from the fact that we cannot observe distant faint sources. As we show in Sect. \ref{Obsd}, the observed spacial distribution of radio pulsars is in agreement with homogeneous distribution. Thus, if $L(P,\chi,B)$ is pulsar luminosity, and $S$ is the receiver sensitivity (radiation is assumed to be isotropic, anisotropy of radiation can be accounted for by properly renormalizing function $L$), one can calculate the impact on the distribution function due to the limited sensitivity:
\begin{align}
    N^{\rm obs} &= \int\limits_{0}^{R_{\rm max}}2\pi l {\rm d}l N \Theta[S - L/4\pi l^2 ] =\\  &= \frac{L(P,\chi,B)}{4 S} N(P,\chi, B), \nonumber
\end{align}
where $\Theta[x]$ is Heaviside function and $R_{\rm max}$ is the characteristic radius of the galactic disk. As most of the pulsars are observed far from the edge of the galactic disk, we assume $R_{\rm max} \rightarrow \infty$.
One can see, that the observed distribution of pulsars is proportional to their intrinsic luminosity. Interestingly, the distribution function does not depend on receiver sensitivity (assuming that it is constant for all pulsars), as it will disappear after normalization. 

Unfortunately, the function $L(P,\chi,B)$ is poorly constrained, as pulsar radio luminosity weakly depends on observed parameters $P$ and $\dot P$. Recent review by \citet{2013IJMPD..2230021B} contains most of the 
proposed models radio luminosity. The dependence of luminosity $L$ on $P$ and $\dot P$ is usually expressed as
\begin{equation}
    L \propto P^{\alpha_1}{\dot P}^{\alpha_2}
\end{equation}
with parameters $\alpha_1$ and $\alpha_2$ used to fit the data. This expression, although widely used, is only observationally motivated. Up to now, there is no physically motivated model describing pulsar luminosity it terms of its intrinsic parameters $P,~\chi, ~{\rm and}~ B$. Early studies \citep[e.g.,][]{1981JApA....2..315V}
tried to find optimal pairs $(\alpha_1,\alpha_2)$ by fitting observed values of luminosities. Such studies give
values $(\alpha_1,\alpha_2) \sim (-0.8, 0.4)$. Later studies \citep[e.g.,][]{2006ApJ...643..332F,
2014MNRAS.439.2893B} take into account selection effects and include $(\alpha_1,\alpha_2)$ as a part of the population synthesis model. Although in principle it should give more accurate values, it introduces two additional free parameters, and makes the synthesis more uncertain. These studies suggest values $(\alpha_1,\alpha_2) \sim (-1.5, 0.5)$.

On the other hand, there are only few theoretical studies of pulsar radio luminosities. For example, counter-alignment model \citep{BGI} predicts
\begin{equation}
L(P,\chi,B) \propto P^{-0.8 \pm 0.2} \cos^{1/2}\chi,
\label{Lbgi}
\end{equation}
while for alignment model there is no such prediction.
Due to such poor constrains on luminosity function, in the majority of the paper we use simplified version
\begin{equation}
L(P,\chi,B) = L(P) \propto P^{-1}.
\label{Pone}
\end{equation}
implying that we can rewrite the distribution function as
\begin{equation}
    N^{\rm obs} = V_{\rm lum}^{\rm vis}(P)N \propto P^{-1} N,
\end{equation}
where 
\begin{equation}
V^{\rm vis}_{\rm lum}(P) \propto P^{-1}.
\label{Vrad}
\end{equation}
This simplification allows us to express results in a compact form. However, our analysis is general and can be easily modified for arbitrary visibility function. We discus the influence of the luminosity function on pulsar statistics as well as the dependence of $L$ on inclination angle and magnetic field in Section~\ref{sect:lum_funct}.

Another important effect is the beaming of radio emission. If pulsar has the inclination angle $\chi$, and the angle between its rotational axis and the direction on observer is $\xi$, the pulsar can be seen only if
\begin{equation}
    |\chi - \xi| < W_0,
\end{equation}
where both angles should lie between $0^\circ$ and $90^\circ$.
It is natural to assume the direction on observer to be randomly distributed. It implies $N(\xi) = \sin\xi$. After that we can determine beaming visibility function from 
\begin{equation}
    V^{\rm vis}_{\rm beam} = \int\limits_{\xi_{\min}}^{\xi_{\max}} \sin\xi{\rm d}\xi,
\end{equation}
where $\xi_{\min} = \max(0,\chi-W_0)$, and $\xi_{\max} = \min(\pi/2,\chi+W_0)$. It is a general expression, and in the limit of small $W_0$ it is consistent with expression (\ref{Vdiag}).

It is necessary to mention that one should be careful when using expression (\ref{Wobs2}). This expression was obtained from observations of \emph{orthogonal} radio pulsars \citep[e.g.,][]{1990ApJ...352..247R}, for which inclination angles are known. The same expression cannot be reliably used for arbitrary angles. As coefficient in (\ref{Wobs2}) implies radio emission coming from the very surface of neutron star, one can expect it to be larger for arbitrary inclined pulsars.

\begin{figure}
\centering
\includegraphics[scale=0.7]{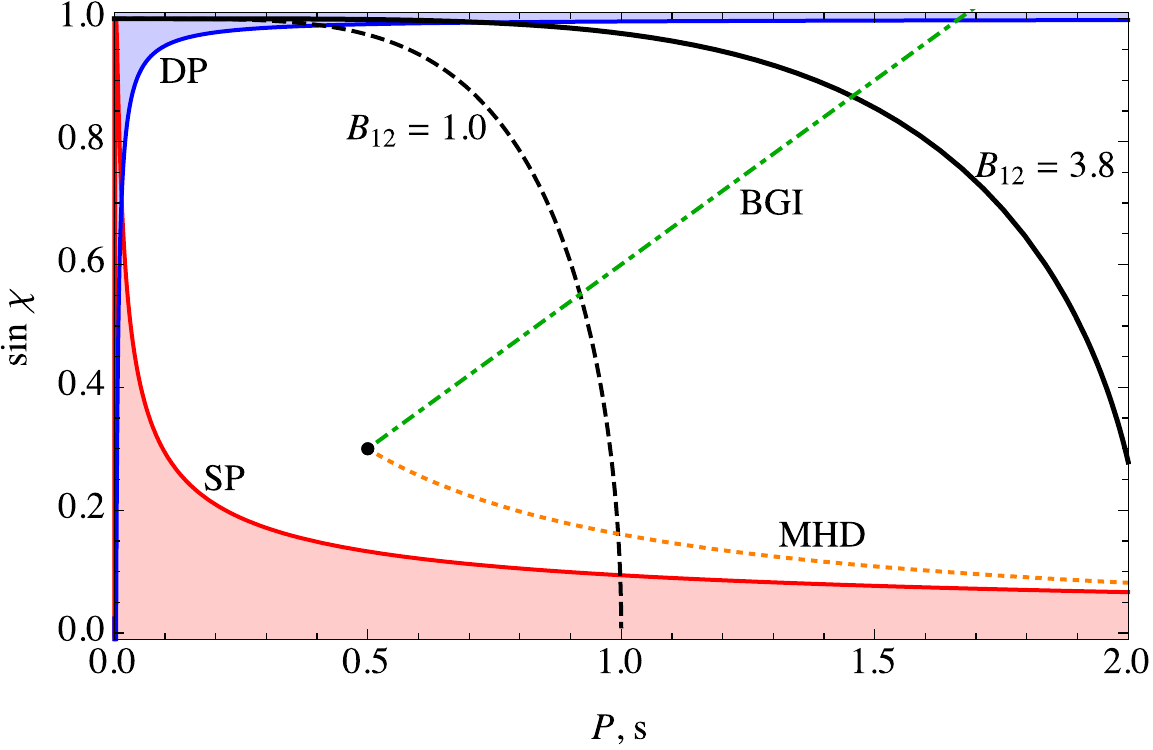}
\caption{
Period-inclination angle diagram. Regions for which it is possible to observe interpulse emission are shaded in blue for double-pole (DP) interpulses, condition (\ref{eq:wj08orth}), and red for single-pole (SP) interpulses, condition (\ref{eq:wj08}). Black lines show the pulsar death line (\ref{deathline}) for $10^{12}$ G magnetic field (dashed line) as well as for Crab-like pulsar with $B_{12} = 3.8$. Green dot-dashed line shows an example of evolution curve according to BGI model (\ref{BGI_P})--(\ref{BGI_chi}), which is straight line. Orange dotted line represents MHD evolution curve according to (\ref{MHD_P})--(\ref{MHD_chi}). Inclination angle of BGI pulsar increases with time, and it inevitably intersects the death line, leading to the suppression in number of DP interpulses.
} 
\label{fig05}
\end{figure}

In addition, the radiation visibility function $V^{\rm vis}_{\rm lum}$ has to take into into 
account the death line which strongly depends upon the inclination angle (see~\citet{2013PhyU...56..164B} 
for more detail). Indeed, as it was already mentioned in the introduction, for given values of  
pulsar period $P$ and magnetic field $B$, the production of particles is suppressed  at angles 
$\chi$ close to $90^{\circ}$, where magnetic dipole moment is nearly orthogonal to the axis of 
rotation. This in turn leads to a decrease in the electric potential drop near the surface of 
neutron star and to suppression of production of secondary particles. E.g., within~\citet{1975ApJ...196...51R}
type model one can write down the following condition for the pair creation~\citep{BGI}
\begin{equation}
\cos\alpha > P^{15/7}B_{12}^{-8/7},
\label{deathline}
\end{equation}
where $B_{12} = B/(10^{12} \, {\rm G})$ and period $P$ is in seconds.
Therefore, neutron stars above and to the right of the extinction lines in Fig.~\ref{fig05} cannot be 
considered as radio pulsars. As it will be discussed below, the death line has to be taken into account 
for orthogonal interpulse pulsars. 

\subsection{Period distribution}
\label{ObsP}

\begin{figure*}
\centering
\includegraphics[scale=0.7]{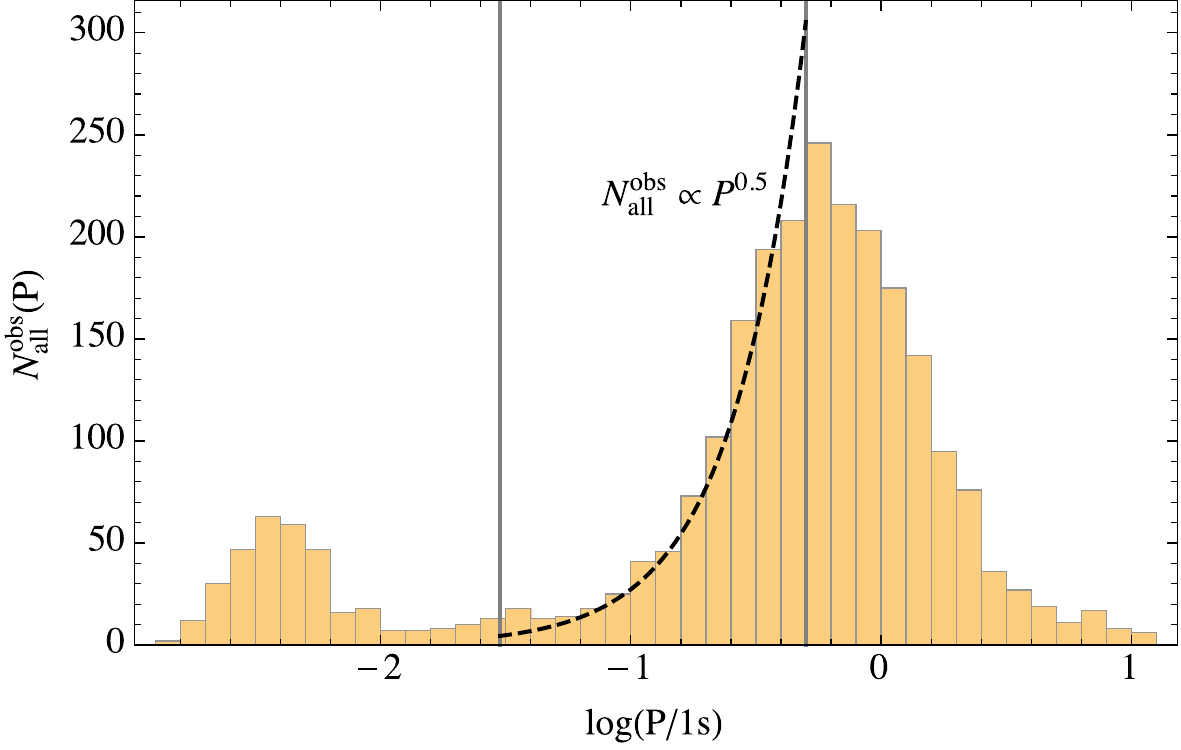}~~~~~\includegraphics[scale=0.7]{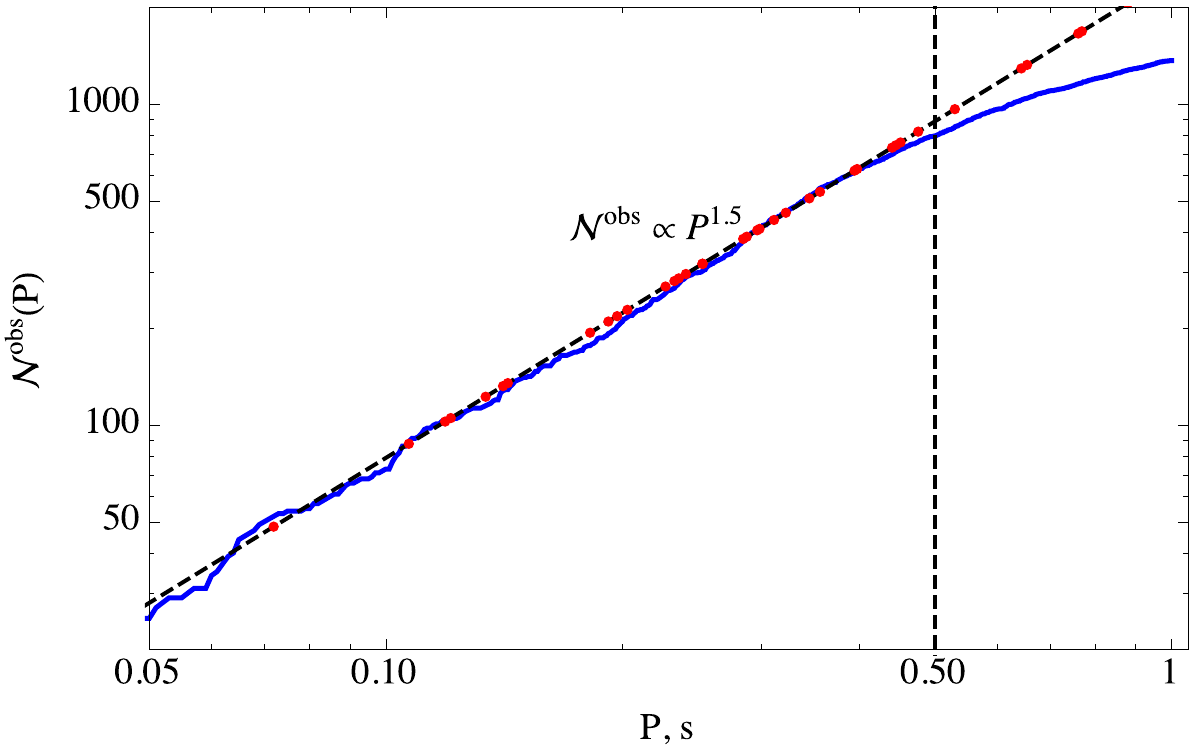}
\caption{
\textbf{[Left panel]:} Period distribution function for all pulsars. The bump to the left corresponds to millisecond pulsars, to the right of it are normal pulsars. In this paper, we consider only pulsars between two vertical lines which have periods $0.03~{\rm s} \le P \le 0.5~{\rm s}$. In this period range, the period distribution is approximately power-law $N^{\rm obs} \propto P^{0.5}$. \textbf{[Right panel]:} Integral period distribution function ${\cal N}^{\rm obs} (P^{\rm obs}<P)$ of normal pulsars. One can see, the distribution function is a power-law until $P\sim0.5~{\rm s}$. Red dots show the locations of pulsars with interpulse emission.
} 
\label{fig01}
\end{figure*}

Finally, let us consider the statistical distribution of the period $P$ which helps 
us evaluate the birth function $Q_{P}(P)$. As is shown on Fig.~\ref{fig01} (left panel), the 
period distribution function $N^{\rm obs}(P)$ contains millisecond branch and normal radio pulsars with 
mean period $P \sim 1$ s. As the evolution of millisecond pulsars differs essentially 
from the evolution of ordinary pulsars \citep[see, e.g.,][]{1998CAS....31.....L}, in what follows we consider the 
pulsars with $P > 0.03$ s only. At first glance, the distribution of ordinary pulsars is 
similar to log-normal one, as is generally assumed for the birth function~\citep{2007PhyU...50.1123P, 2014MNRAS.443.1891G}. 
But, as it is  shown in Fig.~\ref{fig01} (right panel), in reality, for small $P$ distribution function clearly is a power-law
\begin{equation}
N^{\rm obs}(P) \propto P^{0.5}
\label{NsmallPobs}
\end{equation}
until $P\sim 0.5$ s. In what follows we consider \emph{only} the pulsars with $0.03~{\rm s} < P < 0.5~{\rm s}$, and assume (\ref{NsmallPobs}) as their observational distribution function.

Using now the total visibility functions (\ref{Vdiag}) and (\ref{Vrad}), one can conclude 
that $V^{\rm vis} = V^{\rm vis}_{\rm lum} V^{\rm vis}_{\rm beam} \propto P^{-1.5}$. Hence,
the real differential distribution function $N(P)$ for small periods $P$ is to have the form
\begin{equation}
N(P) \propto P^{2}.
\label{NsmallP}
\end{equation}
This power law for small periods is enough for us as the interpulse pulsars have 
rather small periods $P \sim 0.1$--$0.5$ s as well (see red points in Fig. \ref{fig01}). For the same reason, we 
are not going to take into account the evolution of the magnetic field, as the timescale of its evolution is larger than the dynamical age of ordinary pulsars $\tau_{D} \sim P/\dot{P}$.



\subsection{Interpulse pulsars}
\label{sect:interpulse}

\subsubsection{Single/double pole interpulse pulsars}
\label{sect:SD}

As it was already mentioned, interpulse pulsars can provide an insight on the evolution of radio pulsars because they can provide 
additional information about inclination angle $\chi$. Indeed, as it is well-known
~\citep{1977puls.book.....M, 1998CAS....31.....L}, the interpulse appears when we observe either two 
opposite poles (then pulsar will be called DP --- Double Pole), or when we observe the same pole twice
(SP --- Single Pole); 
in the latter case two peaks correspond to 
the double intersection of the hollow-cone directivity pattern. For the DP case the 
inclination angle $\chi$ is close to 90$^{\circ}$, while for SP pulsar this angle is 
close to 0$^{\circ}$. 

It is necessary to underline that sometimes it is rather difficult to make clear distinction 
between single pole and double poles interpulse pulsars. The point is that the procedure of 
determination of inclination angles from polarization characteristics is kind of blurred, 
so some additional arguments are to be used. E.g., one can suppose that for SP-pulsars the 
main pulse/interpulse separation is not equal to 180$^{\circ}$ and is frequency dependent, and 
there is nonzero radio emission between pulses. Accordingly, angular separation of two 
components for DP interpulse pulsars are to be close to 180$^{\circ}$, does not depend on 
the frequency, and there is no radio emission between them.



\subsubsection{Interpulse statistics}
\label{sect:DPSP}

There are several catalogues of interpulse pulsars, most full ones were made by~\citet{2011MNRAS.414.1314M} and by ~\citet{2013ARep...57..833M}.  We collect such pulsars in Table~\ref{table01} 
which includes pulsar names, their periods and period derivatives $P$ and $\dot{P}$, interpulse/mean pulse intensity ratio, and angilar separation between peaks. In addition, 
we mark SP/DP classification taking from~\citet{2011MNRAS.414.1314M, 2013ARep...57..833M}. As one can see, there is 
some disagreement in their interpretation resulting from different approach in 
determination of inclination angle from polarimetric properties. In this work, we do not aim at resolving this disagreement.


\begin{table}
\caption{List of all known interpulse pulsars. [1] -- from \citet{2011MNRAS.414.1314M}, [2] -- from \citet{2013ARep...57..833M}. There is significant disagreement between these studies. We do not aim in resolving this disagreement. Instead, we treat the discrepancy in the classification as the uncertainty in observational constrains (see Table \ref{table02}). 
}  
\centering
\begin{tabular}{|c|c|c|c|c|c|}
  \hline
 Name        &  $P$    & $\dot P$    & IP/MP  &     Sep.      & [1]/[2]  \\
  J          &  [s]    & $10^{-15}$  & ratio  & [$^{\circ}$]  &          \\  
\hline
0534$+$2200 & 0.033  & 423      &0.6            &145    &$-$/$-$    \\
0627$+$0706 & 0.476  & 29.9     &0.2          &180  & DP/DP    \\
0826$+$2637 & 0.53    &  1.7    &0.005        &180  & DP/$-$  \\
0828$-$3417  & 1.85    &  1.0   &0.1         &180  & SP/$-$  \\
0831$-$4406  & 0.312  & 1.3     &0.05           &234    & SP/SP  \\
0834$-$4159  & 0.121  & 4.4     &0.25           &171    & DP/SP     \\
0842$-$4851  & 0.644  & 9.5     &0.14           &180    & DP/DP     \\
0905$-$5127  & 0.346  & 24.9    &0.059          &175    & DP/$-$       \\
0908$-$4913  & 0.107  &15.2     &0.24           &176    & DP/DP     \\
0953$+$0755 & 0.253  & 0.2      &0.012          &210    & SP/SP   \\
1057$-$5226  & 0.197  & 5.8     &0.5            &205    & DP/SP     \\
1107$-$5907  & 0.253  &0.09     &0.2            &191    &SP/DP \\
1126$-$6054  & 0.203  &0.03     &0.1            &174    &DP/DP \\
1244$-$6531  & 1.547  &7.2      &0.3            &145    &DP/SP  \\
1302$-$6350  & 0.047  &2.28     &0.75           &145    &SP/$-$   \\
1413$-$6307  &0.395   &7.434    &0.04           &170    &DP/DP  \\
1424$-$6438  &1.024   &0.24     &0.12           &223    &SP/SP  \\
1549$-$4848  &0.288   &14.1     &0.03           &180    &DP/DP  \\
1611$-$5209  &0.182   &5.2      &0.1            &177    &DP/$-$         \\
1613$-$5234  &0.655   &6.6      &0.28           &175    &DP/$-$          \\
1627$-$4706  &0.141   &1.7      &0.13           &171    &DP/SP \\
1637$-$4553  &0.119   &3.2      &0.1            &173    &DP/DP \\
1637$-$4450  &0.253   &0.58     &0.26           &256    &SP/SP  \\
1705$-$1906  &0.299   &4.1      &0.15           &180    &DP/DP  \\
1713$-$3844  &1.600   &177.4    &0.25           &181    &DP/$-$      \\
1722$-$3712  &0.236   &10.9     &0.15           &180    &DP/DP   \\
1737$-$3555  &0.398   &6.12     &0.04           &180    &DP/SP  \\
1739$-$2903  &0.323   &7.9      &0.4            &180    &DP/DP \\
1806$-$1920  &0.880   &0.017    &1.0            &136    &SP/SP \\
1808$-$1726  &0.241   &0.012    &0.5            &223    &SP/SP \\
1825$-$0935  &0.769   &52.3     &0.05           &185    &$-$/SP       \\
1828$-$1101  &0.072   &14.8     &0.3            &180    &DP/$-$   \\
1842$+$0358 &0.233   &0.81      &0.23           &175    &DP/$-$  \\
1843$-$0702  &0.192   &2.1      &0.44           &180    &DP/$-$  \\
1849$+$0409 &0.761   &21.6      &0.5            &181    &DP/$-$    \\
1851$+$0418 &0.285   &1.1       &0.2            &200    &SP/SP \\
1852$-$0118  &0.452   &1.8      &0.4            &144    &SP/SP \\
1903$+$0925 &0.357   &36.9      &0.19           &240    &SP/SP  \\
1913$+$0832 &0.134   &4.6       &0.6            &180    &DP/$-$      \\
1915$+$1410 &0.297   &0.05      &0.21           &186    &DP/$-$      \\
1932$+$1059 &0.227   &1.2       &0.018          &170    &DP/SP  \\
1946$+$1805 &0.441   &0.02      &0.005          &175    &SP/SP \\
2032$+$4127 &0.143   &20.1      &0.18           &195    &DP/SP  \\
2047$+$5029 &0.446   &4.2       &0.6            &175    &DP/$-$    \\  
  \hline
\end{tabular} 
\label{table01} 
\end{table}

It is necessary to stress that one of the main features of interpulse pulsars 
is their rather small periods $P$ against total population as presented on Fig.~\ref{fig01}. 
Accordingly, their dynamical ages $\tau_{\rm D} \approx P/2{\dot P}$ are much less than for 
the most of radio pulsar ($\sim 1$--$10$ Myrs). Besides, as is shown in Table~\ref{table02},  
the number of interpulse pulsars in the period range \mbox{$0.03$ s $< P < 0.5$ s} is much 
larger than outside of this range $P>0.5$ s. Thus, by considering this period range and using distribution function of all pulsars $N^{\rm obs}\propto P^{0.5}$, we can describe most of interpulse pulsars with good accuracy.


\begin{table}
\caption{Number of interpulse pulsars. The lower values correspond to certain classification 
(the same determination in [1] and [2]) with high enough interpulse-main pulse 
intensity ratio IP/MP $> 0.1$.}
\centering
\begin{tabular}{|r|c|c|}
    \hline
        &  $0.03 \div 0.5$ s & $> \, $0.5 s     \\
\hline
${\cal N}_{\rm SP}$, observed     &       $4 \div 10$    &  $2 \div 3$    \\  
${\cal N}_{\rm  DP}$, observed    &      $10 \div 24$    &  $3 \div 5$   \\  
  \hline
\end{tabular}
\label{table02} 
\end{table}

\subsubsection{Visibility function}
\label{sect:vis_interpulses}

For interpulse pulsars it is necessary to make a correction to the beaming visibility function. For almost orthogonal double-pole interpulses, the condition to see two oppositely directed poles has a form
\begin{equation}
\label{eq:wj08orth}
    \pi-\chi-\xi < W_0,
\end{equation}
which means that the visibility function
\begin{equation}
\label{eq:vis_DP}
    V^{\rm vis,~DP}_{\rm beam} = \int\limits_{\xi_{\rm min}^{\rm DP}}^{\pi/2} \sin\xi{\rm d}\xi,\quad \xi_{\rm min}^{\rm DP} = \min(\pi/2,\pi-W_0-\chi).
\end{equation}

\begin{figure}
\centering
\includegraphics[scale=0.9]{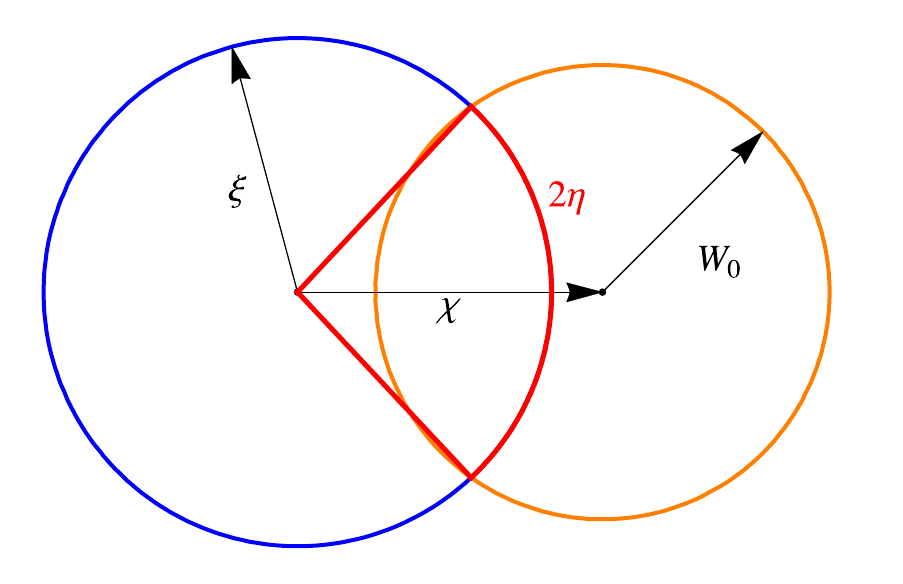}
\caption{
Illustration for single pole interpulse selection criterion. The orange circle represents the boundary of emission region. In the reference frame rotating with the star, the blue circle corresponds to the `trajectory' of the line of sight. As for single-pole interpulses all angles $\xi$, $\chi$, and $W_0$ must be small, we can assume that all circles are in the same plane and get a condition (\ref{eq:circle}) for the visible fraction of a period (red) to be larger than $2\eta$. 
} 
\label{fig:circle}
\end{figure}

For single pole interpulses, the condition to see the same pole twice is \citep{2008MNRAS.387.1755W}
\begin{equation}
\label{eq:wj08}
    \chi + \xi < W_0,
\end{equation}
implying the visibility function
\begin{equation}
\label{eq:vis_SP}
    V^{\rm vis,~SP}_{\rm beam} = \int\limits_{0}^{\xi_{\rm max}^{\rm SP}} \sin\xi{\rm d}\xi,\quad \xi_{\rm max}^{\rm SP} = \max(0,W_0-\chi).
\end{equation}
However, equation (\ref{eq:wj08}) underestimates the fraction of single pole interpulses. Equation (\ref{eq:wj08}) implies that the observer can see the emission region over the whole rotation period. But given 
that the angular separation between the main pulse and interpulse for SP interpulses is often less that 180$^{\circ}$ (see Table \ref{table02}), we can come out with the following {\it necessary} condition 
for single pole interpulse:
\begin{equation}
\label{eq:circle}
\xi^2 + \chi^2 - 2\xi \chi\cos\eta  \le W_0^2,
\end{equation}
where $\eta$ represents the fraction of the period during which observer can see the emission region 
(see Figure \ref{fig:circle} for clarification). If $\eta = \eta_{\rm max} = 180^\circ$ equation (\ref{eq:circle}) gives the same constrain as equation (\ref{eq:wj08}). However, from Table \ref{table01} 
one can see that the separation between the main pulse and interpulse for SP pulsars can be as low as 
$136^\circ$ implying $\eta_{\rm min} = 68^\circ$. In what follows, we estimate the number of single pole interpulses for both $\eta_{\rm min}$ and $\eta_{\rm max}$, which gives us upper and lower boundaries for interpulse fractions.

It is worth noting that the conditions (\ref{eq:vis_SP}) and (\ref{eq:vis_DP}) do not depend on whether the emission comes from core or conal component. The geometry of emission region is parametrized by a single parameter $W_0$. By changing this parameter one can consider core and conal components separately.


\section{Evolution theories}
\label{sect:torques}
 
\subsection{Current losses}

As was mentioned above, one of the ways to understand pulsar braking mechanisms is to analyse the inclination
angle evolution. In present paper we consider two magnetospheric theories, both predicting simple analytical
expressions for time evolution of period $P$ and inclination angle $\chi$. The first one is the numerical
force-free/MHD model~\citep{2006ApJ...648L..51S, 2014MNRAS.441.1879P} predicting evolution towards $0^{\circ}$.
Another one is related to quasi-analytical model elaborated by~\citet{1984Ap&SS.102..301B, BGI} and predicts counter-alignment. In both cases we do not include into consideration magnetic field evolution since most of
interpulse pulsars have dynamical ages smaller than characteristic timescales of magnetic field evolution.  

The braking of the neutron star rotation results from impact of the torque ${\bf K}$ due 
to longitudinal currents $j_{\parallel}$ circulating in the pulsar magnetosphere; for 
zero longitudinal current the magneto-dipole radiation of a star is fully screened by radiation of the pulsar magnetosphere~\citep{BGI, MPS99}. General expressions connecting 
the time evolution of the angular velocity $\Omega$ and inclination angle $\chi$ can be
parametrized as~\citep{BGI, 2014MNRAS.441.1879P}
\begin{eqnarray}
I_{\rm r} \, \dot{\Omega}
& = & K_{\parallel}\cos{\chi}+K_{\perp}\sin{\chi},
\label{18'} \\
I_{\rm r}\Omega \, {\dot\chi}
& = & K_{\perp}\cos{\chi}-K_{\parallel}\sin{\chi},
\label{19'}
\end{eqnarray}
where $I_{r} \propto M R^2$ is the neutron star momentum of inertia and we introduce two components of the 
torque ${\bf K}$ parallel and perpendicular to the magnetic dipole ${\bf m}$. 

It is convenient to describe these values by dimensionless current 
$i \approx j_{\parallel}/j_{\rm GJ}$ by separating it into symmetric 
part $i_{\rm s}$ (which has the same sign in the northern and 
southern parts of the polar cap), and antisymmetric part $i_{\rm a}$ 
(which reverts sign on the polar cap).  Here and below we apply 
normalization to the `local' Goldreich-Julian current 
density, $j_{\rm GJ} = |{\bf \Omega}\cdot{\bf B}|/2\pi$ (with scalar product). For dipole 
magnetic field and small angles $\theta - \chi \sim (\Omega R/c)^{1/2}$ we have
\begin{equation}
\label{jgj}
j_{\rm GJ}(r_{m}, \varphi_{m}) \approx \frac{\Omega B_{0}}{2 \pi} 
\left(\cos\chi + \frac{3}{2} \, \frac{r_{m}\sin\varphi_{m}}{R}\sin\chi\right).
\end{equation}
Here $B_0$ is magnetic field on the neutron star magnetic pole, $R$ is neutron star radius, and $r_{m}$ and $\varphi_{m}$ are polar co-ordianates in the magnetic polar cap. 
As a result, one can write down
\begin{eqnarray}
i_{\rm s} & = & i_{\rm s}^{A} \cos\chi,
\label{16new} \\
i_{\rm a} & = & i_{\rm a}^{A} \sin\chi,
\label{17new}
\end{eqnarray}
where the amplitude values
\begin{eqnarray}
i_{\rm s}^{A} & = &  \frac{2(I_{+} + I_{-})}{\Omega B_{0}R_{0}^2 \cos\chi} ,
\label{16nn} \\
i_{\rm a}^{A} & = &  \frac{\pi R(I_{+} - I_{-})}{\Omega B_{0}R_{0}^3\sin\chi} 
\label{17nn}
\end{eqnarray}
can be determined by the currents through the northern and southern parts of the polar cap
\begin{eqnarray}
I_{+} & = &  \int_{0}^{R_{0}}\int_{0}^{\pi} j_{\parallel} r_{m} {\rm d}r_{m} {\rm d}\varphi_{m},
\label{16nn} \\
I_{-} & = &  \int_{0}^{R_{0}}\int_{\pi}^{2\pi} j_{\parallel} r_{m}  {\rm d}r_{m} {\rm d}\varphi_{m}.
\label{17nn}
\end{eqnarray} 
Here $R_{0} \approx (\Omega R/c)^{1/2}R$ is the polar cap radius.
For \mbox{$j_{\parallel} = j_{\rm GJ}$} we have $i_{\rm s}^{A}= i_{\rm a}^{A} = 1$.

As one easily check, $K_{\parallel} \propto i_{\rm s}$, and  $K_{\perp} \propto i_{\rm a}$. 
In particular, the direct action of the Amp\`ere force on the star by surface currents (which 
are close to the longitudinal electric currents circulating in the pulsar magnetosphere) can be
written as~\citep{1984Ap&SS.102..301B}
\begin{eqnarray}
K_{\parallel}^{\rm sur} & = &
- c_{\parallel} \frac{B_{0}^{2}\Omega^{3}R^{6}}{c^{3}} i_{\rm s},
\label{16'} \\
K_{\perp}^{\rm sur} & = & - c_{\perp} \frac{B_{0}^{2}\Omega^{3}R^{6}}{c^{3}}
\left(\frac{\Omega R}{c}\right)i_{\rm a}.
\label{17'}
\end{eqnarray}
Here  the coefficients $c_{\parallel}$ and $c_{\perp}$ are factors of order unity which depend on the profile 
of the longitudinal current and polar cap form. As we see, for `local' Goldreich-Julian current 
$i_{\rm s} \approx i_{\rm a} \approx 1$ relations \eqref{16'} and \eqref{17'} imply that
\begin{eqnarray}
K_{\perp}^{\rm sur} \approx \left(\frac{\Omega R}{c}\right) K_{\parallel}^{\rm sur},
\label{iais}
\end{eqnarray}
so that $K_{\perp}^{\rm sur} \ll K_{\parallel}^{\rm sur}$. Below we also assume (as was not done up to now) 
that the additional contribution for $K_{\perp}$ can give the magnetosphere itself, more precisely, the 
mismatch between magneto-dipole radiation from magnetized star and radiation from the magnetosphere (which 
exactly compensate themselves in case of zero longitudinal current). Here we write down $K_{\perp}^{\rm mag}$ 
in general form as
\begin{eqnarray}
K_{\perp}^{\rm mag} & = & - A \, \frac{B_{0}^{2}\Omega^{3}R^{6}}{c^{3}} \, i_{\rm a}
\label{17''}
\end{eqnarray}
trying to evaluate the dimensionless constant $A$ later from the results of numerical simulations.

Introducing now amplitude values $K_{\parallel}^{A} = K_{\parallel}(0)$ and $K_{\perp}^{A} = K_{\perp}(\pi/2)$,
we
finally obtain
\begin{eqnarray}
I_{\rm r}\dot{\Omega}
& = & K_{\parallel}^{A} + (K_{\perp}^{A}-K_{\parallel}^{A})\sin^2\chi,
\label{8'} \\
I_{\rm r}\Omega {\dot\chi}
& = & (K_{\perp}^{A}-K_{\parallel}^{A})\sin\chi\cos\chi,
\label{9'}
\end{eqnarray}
As both expressions contain the same factor $(K_{\perp}^{A}-K_{\parallel}^{A})$, 
one can conclude that the sign of $\dot\chi$ is associated with $\chi$-dependence of the 
energy losses \citep{2013PhyU...56..164B}. In other words, inclination angle $\chi$ evolves 
towards $90^{\circ}$ (counter-alignment) if total energy losses decrease with increasing 
inclination angles, and towards $0^{\circ}$ (alignment) if they increase with inclination angle.

\subsection{Two braking models}

\subsubsection{Force-free/MHD model (alignment)}
\label{sect:torqueMHD}
 
According to force-free/MHD model worked out on the basis of recent numerical simulations~\citep{2014MNRAS.441.1879P}, 
the rotation braking and the inclination angle evolution can be approximately defined 
as
\begin{eqnarray}
\dot\Omega & \approx & -\frac{1}{4} \frac{B_{0}^{2}R^{6}\Omega^{3}}{I_{r} c^{3}}(1+\sin^{2}\chi),
\label{MHD_P}\\ 
\dot\chi & \approx & -\frac{1}{4} \frac{B_{0}^{2}R^{6}\Omega^{2}}{I_{r} c^{3}} \sin\chi \cos\chi.
\label{MHD_chi}
\end{eqnarray}
Accordingly, the total magnetospheric losses are
\begin{equation}
W_{\rm tot}^{{\rm MHD}} \approx \frac{1}{4} \frac{B_{0}^{2}\Omega^{4}R^{6}}{c^{3}}(1+\sin^{2}\chi),
\label{totMHD}
\end{equation}
i.e., they increase along with inclination angle $\chi$. Evolution law 
(\ref{MHD_P})--(\ref{MHD_chi}) has an `integral of motion'  
\begin{equation}
\label{IMHD}
I^{\rm MHD} = \frac{P \sin \chi}{\cos^{2} \chi}, 
\end{equation}
which will be used in what follows. As we see, in this model the inclination angle $\chi$ 
evolves to 0$^{\circ}$. 

It is necessary to point out that, according to (\ref{iais}), this case can be realised either for strong 
enough anti-symmetric current $i_{\rm a}^{\rm A} \sim (\Omega R/c)^{-1}$, or for large enough contribution 
of the  magnetospheric torque (\ref{17''}). But as one can easily find by analysing analytical asymptotic 
behavior of quasi-radial MHD flows (see, e.g.,~\citealt{PTS16}), MHD solution (\ref{totMHD}) corresponds 
to insufficient value $i_{\rm a}^{\rm A} \sim (\Omega R/c)^{-1/2}$. Remembering that dimensionless current 
$i_{\rm a}$ was normalised to `local' Goldreich-Julian current $j_{\rm GJ}^{\rm loc}$, we see that the 
total current circulating in the magnetosphere of the orthogonal rotator is similar to axisymmetric 
case~\citep{2010ApJ...715.1282B}. It is not surprising because just this total electric current is necessary for the toroidal 
magnetic field on the light cylinder to coincide with electric one. This antisymmetric current for 
ordinary pulsars with $P \sim 1$ s is to be $10^2$ times larger than local Goldreich-Julian current. 
The possibility for the longitudinal current to be much larger than Goldreich-Julian one was recently 
discussed by~\citet{2013MNRAS.429...20T}. 

Thus, one can conclude that to explain MHD energy losses it is necessary to suppose the existence
of magnetospheric losses with 
\begin{equation}
A \approx  2 \left(\frac{\Omega  R}{c}\right)^{1/2}.
\label{A}
\end{equation}
Resulting from large enough anti-symmetric currents \mbox{$i_{\rm a} \gg 1$,} it gives the necessary 
contribution to total energy losses.

\subsubsection{BGI model (counter-alignment)} 
\label{sect:model}
 
Analytical theory of pulsar magnetosphere formulated by~\citet{1984Ap&SS.102..301B, BGI} is based on
three key assumptions:
\begin{itemize}
\item 
longitudinal current $j_{\rm s}$ circulating in pulsar magnetosphere does not exceed 
the local one $j_{\rm GJ} \approx \Omega B_{0} \cos\chi/2 \pi$; its value
is determined by potential drop in the inner gap $V$
\begin{equation}
i_{\rm s}^{A} \approx \frac{1}{2}\left(\frac{V}{V_{\rm max}}\right)^{1/2},  
\end{equation}
where $V_{\rm max} = (\Omega R/c)^{2}R B_{0}$ is the maximum potential drop;
\item
potential drop $V$ is determined by ~\citet{1975ApJ...196...51R} model;
\item
magnetospheric contribution $K_{\perp}^{\rm mag}$ (\ref{17''}) was neglected; as now becomes clear from
(\ref{17''}) and (\ref{A}), this assumption is indeed correct for small anti-symmetric longitudinal 
current $i_{\rm a} \sim 1$ which was also postulated.
\end{itemize}
As a result, this model provides the following evolution law for $\cos\chi > (\Omega R/c)^{-1}$
\begin{eqnarray}
\dot{P}_{-15} & = & Q_{\rm BGI}\frac{B^{2}_{12}}{P} \cos^{2}\chi,  
\label{BGI_P} \\
 \dot{\chi} & = & Q_{\rm BGI} \frac{B^{2}_{12}}{P} \sin\chi \cos\chi,
\label{BGI_chi}
\end{eqnarray}
where again $B_{12} = B_{0}/(10^{12} \, {\rm G})$ and $\dot{P}_{-15} = \dot{P}/10^{-15}$ are 
normalized magnetic field and period derivative respectively, and $P$ is given in seconds. 
As to the main dimensionless parameter of this theory $Q_{\rm BGI} \approx j/j_{\rm GJ}$, for $Q_{\rm BGI} < 1$ 
it can be defined as~\citep{1984Ap&SS.102..301B} 
\begin{equation}
Q_{\rm BGI} = P^{15/14}B^{-4/7}_{12} \cos^{2d-2}\chi,
 \end{equation}
where $d \approx 0.75$. For $Q_{\rm BGI} > 1$ one has to put $Q_{\rm BGI} = 1$.

As a result, for $\chi \neq 90^{\circ}$ the Euler equation predicts the conservation of the following 
invariant:
 \begin{equation}
I^{\rm BGI} = \frac{P}{\sin\chi}.
\label{IBGI}
\end{equation}
Thus, within this model the polar angle $\chi$ shall increase with time. Accordingly, 
total energy losses decrease along with inclination angle $\chi$ increase
\begin{equation}
W_{\rm tot}^{({\rm BGI})} 
\approx i_{\rm s}^{\rm A}\frac{B_{0}^{2}\Omega^{4}R^{6}}{c^{3}}\cos^{2}\chi.
\end{equation}
Finally, for $Q_{\rm BGI} < 1$ radio luminosity $L$ can be presented as
$L = \alpha W_{\rm part}$, where $W_{\rm part} = Q_{\rm BGI}^2 W_{\rm tot}$
is the particle energy flux and $\alpha \sim 10^{-6}$ is the transformation
coefficient. It gives
\begin{equation}
L^{\rm BGI} \propto P^{-0.8} \cos^{1/2}\chi.
\label{LBGI}
\end{equation}
For $\cos \chi \sim 1$ we return to the evaluation similar to (\ref{Vrad}).

\section{Predictions vs observations} 
\label{sect:solution}
\subsection{General assumptions}

\subsubsection{Preliminary remarks}
\label{sect:Remarks}

To clarify the mechanism of radio pulsar braking we determine the number 
of radio pulsars having such angles $\chi$, so they can be observed as interpulse
pulsars. As their period distribution depends directly on their evolution, it gives us the
possibility to recognize the direction of the inclination angle evolution as well. For this reason, 
we consider the pulsars with 0.03 s $<P<$ 0.5 s for two evolutionary scenario (\ref{MHD_P})--(\ref{MHD_chi}) 
and (\ref{BGI_P})--(\ref{BGI_chi}) using kinetic equation method. 
There are two important points to be mentioned.

First, with no regard to the smallness of period $P$, for interpulse pulsars the death line shall 
be taken into consideration for the orthogonal case for BGI model. As shown on Fig.~\ref{fig05}, for inclination 
angle $\chi$ close to $90^{\circ}$ the death line on the $P$--$\sin \chi$ diagram is located at 
small enough periods $P < 1$ s. Moreover, in this case the shape of the region within the polar 
cap where most of the radio emission is produced is not well understood. 

Indeed, 
it is impossible to create pairs both near the 
line where the Goldrech-Julian charge density changes sign preventing longitudinal electric 
field to be large enough. As a result, the geometrical visibility function $V_{\rm beam}^{\rm vis}$
cannot be determined with sufficient accuracy. 

On the other hand, numerical kinetic simulations of nearly MHD magnetospheres \citep{2015ApJ...801L..19P} show abundant pair production for large inclination angles. 
Therefore, for BGI model we consider only SP 
interpulses for which the death line cannot play important role, while for MHD model we consider DP interpulses as well.


Second, we assume, that in the pulsar birth function $Q(P,\chi,B,\xi)$ all arguments are independent 
of each other:
\begin{equation}
    Q(P,\chi,B,\xi) = Q_P(P)Q_{\chi}(\chi)Q_{B}(B)\sin\xi.
\end{equation}
As we already stressed out, evolution of magnetic field is not important for short dynamical ages. This 
allows us to obtain an exact solution of kinetic equation with period distribution, which does not depend 
on magnetic field birth function.

\subsubsection{Initial periods and inclination angles}
\label{Initial}

As was already stressed out, visible distribution of radio pulsars strongly depends on their 
initial periods $P$ and inclination angles $\chi$. Repeated attempts were made to 
determine the birth function $Q_{P}(P)$~\citep{1985MNRAS.213..613L, 2012Ap&SS.341..457P}, but so far, this function 
remains unknown. The new point of our paper is that we use here the direct observational 
scaling ${ N}^{\rm obs}(P) \propto P^{0.5}$ shown on Fig.~\ref{fig01}. 
Being valid for short periods $P < 0.5$ s, this distribution is to describe interpulse pulsars 
with high enough precision.

As for the birth function $Q_{\chi}$ describing the distribution on initial inclination angles
$\chi$, we consider two possibilities, namely $Q_{\chi} = \sin\chi$ and $Q_{\chi} = 2/\pi$. The first 
one corresponds to the random orientation of the magnetic axis with respect to rotation one which is 
more reasonable at first glance. But as we will see,  observational evaluation of the real
$\chi$-distribution $N(\chi)$ (\ref{const}) can correspond to the homogeneous distribution  
$Q_{\chi} = 2/\pi$ as well.

\subsubsection{Comparison with observations}

To compare the predictions of evolutionary scenarios with observations it is not sufficient to 
know the distribution function of radio pulsars $N(P, \chi)$ because it is necessary to include 
into consideration the visibility functions $V^{\rm vis}$ (see Sect.~2.1). In particular, the 
visible distribution of the SP interpulse pulsars is to be written as 
\begin{equation}
 N^{\rm obs}(P) = \int_{0}^{W_{r}(P)}{\rm d}\chi \, V^{\rm vis}(P, \chi) \, N(P, \chi).
\label{norm1}
\end{equation}
Relation (\ref{norm1}) helps us to normalize the observed distribution function as well. 
We normalize the distribution function by the total number of observed pulsars in the period rage 
0.03 s $< P < $ 0.5 s
\begin{equation}
 {\cal N}_{\rm tot} = \int_{0.03}^{0.5}{\rm d}P \int_{0}^{\pi/2}{\rm d}\chi V^{\rm vis}(P, \chi) \, N(P, \chi) .
\label{norm2}
\end{equation}
Observationally, we know that ${\cal N}_{\rm tot} = 796$, and in what follows we use this number to 
normalize the distribution function. 

 
\subsection{Population synthesis -- kinetic equation}
\label{sect:synthesisKE}

In this subsection we describe our approach of using the kinetic equation 
\begin{equation}
\label{eq:main}
\frac{\partial }{\partial P} (\dot{P} \, N) + \frac{\partial}{\partial \chi} (\dot{ \chi} \, N) = Q
\end{equation}
to obtain the real distribution function $N(P, \chi)$ of radio pulsars. Here the values $\dot{P}(P, \chi)$ 
and $\dot{ \chi}(P, \chi)$ are to be taken from the given model. Accordingly, $Q(P, \chi)$ is the 
birth function depending both on the inclination angle $\chi$  and initial period $P$. Here for simplicity 
we put $Q(P, \chi) = Q_{P}(P) Q_{\chi}(\chi)$. Certainly, we also assumed that the observable distribution is time-independent 
due to very small dynamical life time $\tau_{\rm D} = P/{\dot P} \sim 10$ Myr in comparison with Galactic age.
Finally, we do not consider magnetic fields in this Section, and discuss their impact in Section \ref{sect:mag_fields}. 
 
Due to the existence of integrals of motion, kinetic equation can be easily solved. Then, adding the 
visibility functions $V^{\rm vis} = V_{\rm lum}^{\rm vis}V_{\rm beam}^{\rm vis}$ discussed in
Sect. \ref{sect:vis_f} one can determine the number of observed pulsars and compare it with observations. 

As a result, for force-free/MHD model (\ref{MHD_P})--(\ref{MHD_chi}) the kinetic equation has a form
 \begin{equation}
\frac{\partial}{\partial P} \left[\frac{N}{P}(1 + \sin^{2} \chi)\right] 
- \frac{\partial}{\partial \chi}\left[\frac{N}{P^{2}} \sin{\chi} \cos{\chi}\right]
=  K Q,
 \end{equation}
where $K = I_{r}c^{3}/(\pi^{2}B^{2}R^{6})$. In what follows we neglect this factor as it disappears after normalisation (see also Section \ref{sect:mag_fields}).
Using now expression (\ref{IMHD}) for the `integral of motion' $I^{\rm MHD}$ we obtain a solution 
which is valid for arbitrary $Q_P$ and $Q_\chi$:
\begin{equation}
N^{{\rm MHD}}= \frac{P^2}{\cos^3 \chi} \int\limits_\chi^{\pi/2} 
\frac{\cos^2 x}{\sin x}Q_\chi(x)Q_P\left(P \frac{\sin \chi}{\cos^2 \chi}\frac{\cos^2 x}{\sin x} \right) {\rm d}x.
\label{kin1}
 \end{equation}
 Note that one needs to normalize this solution according to (\ref{norm2}). Assuming $N(P) \propto P^2$ (\ref{NsmallP}), one can obtain birth function $Q_P = {\rm const}$.
  As a result, for $Q_{\chi} = 2/\pi$, the solution has a simple form:
\begin{equation}
N^{{\rm MHD}}(P, \chi) = - \,
\frac{2 \log \tan(\chi/2) + 2\cos{\chi}}{\cos^{3} \chi} \, P^{2},
\label{MHD1}
\end{equation}
Accordingly, for $Q_{\chi} = \sin\chi$ we have
\begin{equation}
N^{{\rm MHD}}(P, \chi) =  \frac{\pi/2 - \chi - \sin \chi \cos \chi}{\cos^3 \chi} P^2.
 \label{MHD2}
\end{equation}

As to BGI model (\ref{BGI_P})--(\ref{BGI_chi}), the kinetic equation has the form:
\begin{equation}
\frac{\partial}{\partial P} \left[N \cos^{2d} {\chi}\right] 
+ \frac{\partial}{\partial \chi} \left[\frac{N}{P} \sin \chi \cos^{2d-1} \chi\right] 
= Q_{P}(P) Q_{\chi}(\chi). 
\end{equation}
Using again the `integral of motion' $I^{\rm BGI}$ (\ref{IBGI}), we obtain
\begin{equation}
N^{{\rm BGI}}(P, \chi) = \frac{ P}{\sin^{2}{ \chi} \cos^{2d-1} \chi} \int\limits_0^\chi  Q_\chi(x) Q_P\left( P \frac{\sin x}{\sin \chi}\right) \sin x {\rm d} x,
\end{equation}
which again should be propely normalized.
As $N(P) \propto P^2$, one can conclude that $Q_{P}(p)$ is to be linear function of $p$. 
As a result, for homogeneous angular birth function $Q_{\chi} = 2/\pi$, the solution looks like
\begin{equation}
 N^{{\rm BGI}}(P, \chi) = \frac{\chi - \sin\chi \cos\chi}{\sin^{3}{ \chi} \cos^{2d-1} \chi} P^2.
 \label{BGI1}
 \end{equation}
 Accordingly, applying the `random' angular birth function $Q_{\chi} = \sin\chi$ we obtain
 \begin{equation}
 N^{{\rm BGI}}(P, \chi) =  \frac{2 + \cos^3\chi - 3 \, \cos\chi}{\sin^{3} \chi \cos^{2d-1} \chi} P^2.
  \label{BGI2}
 \end{equation}
 
\begin{figure}
\includegraphics[scale=0.57]{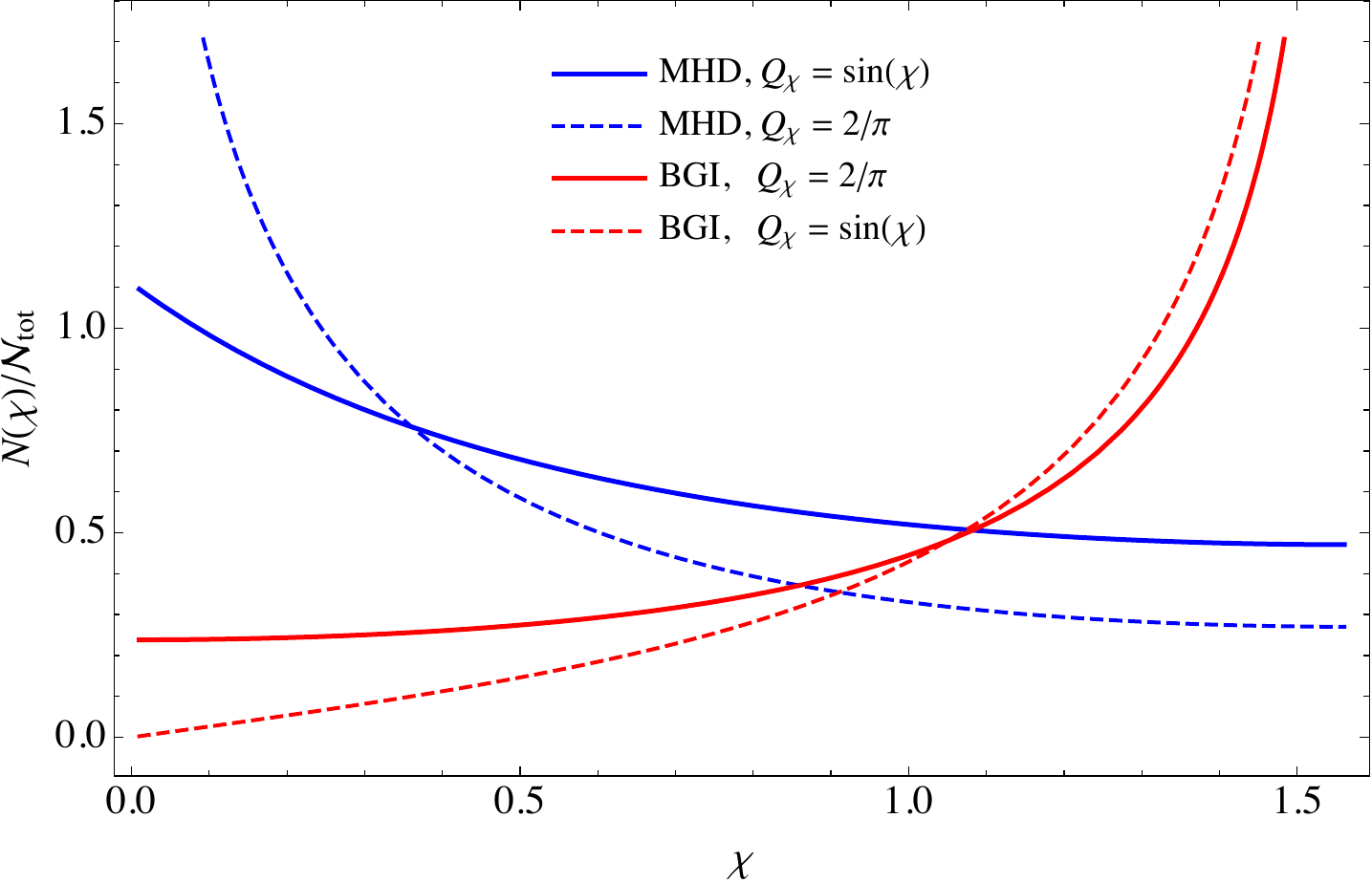}
\caption{
The comparison of angular distribution functions $N(\chi)$ for different evolution models (\ref{MHD1}), (\ref{MHD2}), (\ref{BGI1}), and (\ref{BGI2}). Models described with dashed lines do not have a finite value in the limit of $\chi \rightarrow 0$, and thus are in contradiction with observations. On the other hand, models described with solid lines are in agreement with observations. This shows that MHD and BGI models require different birth function $Q_\chi$ to be consistent with observations.
} 
\label{fig:chi-dist-all}
\end{figure}

 We present angular distribution functions for different models in Figure \ref{fig:chi-dist-all}. One can easily notice, that for MHD model with uniform angular birth function (blue, dashed line), the number of pulsars with small angles is very large, while for BGI model and sinusoidal angular distribution function (red dashed line), the fraction of pulsar with small angle is close to zero. However, observationally one has $N(\chi) \approx {\rm const}$ at small angles. This implies that the models described with dashed lines in Figure \ref{fig:chi-dist-all} are inaccurate. On the other hand, models described with solid lines both have finite limit at $\chi \rightarrow 0$, and thus are in agreement with observations. One should also remember that these distribution functions should be corrected with visibility function in order to obtain observed distribution function.
 
 \subsection{Dependence on the magnetic fields}
 \label{sect:mag_fields}
One can immediately see from equation (\ref{eq:main}) that
\begin{align}
N& \propto Q_\xi(\xi) Q_B(B) \, B^{-2} ~~~~~~ &{\rm (MHD)}; \label{eq:n_b_1}\\
N& \propto Q_\xi(\xi) Q_B(B) \, B^{-10/7} ~~~ &{\rm (BGI)}.\label{eq:n_b_2}
\end{align}
It turns out that the observed distribution function $N^{\rm obs} (P,\chi)$ does not depend on the form of birth function $Q_B(B)$. To show that, we consider MHD case only, but the same conclusion remains true for BGI model as well since the only difference is in the power of $B$ in the denominator (\ref{eq:n_b_1})-(\ref{eq:n_b_2}). The observed distribution function $N^{\rm obs} (P,\chi)$ is given by
\begin{align}
N^{\rm obs} (P,\chi) &= \int\limits_0^\infty {\rm d} B\int\limits_0^{\pi/2} {\rm d} \xi V_{\rm vis} (P,\chi,B,\xi) N(P,\chi,B,\xi),\propto \nonumber\\
&\propto\int\limits_0^\infty {\rm d} B V^{\rm vis}_{\rm lum}(P,\chi,B) Q_B(B)/B^2.
\end{align}
Then, assuming $V_{\rm lum}^{\rm vis} \propto \dot P^{\alpha_2} \propto B^{2\alpha_2}$, we get $N^{\rm obs} \propto \int {\rm d}B B^{2 \alpha_2-2} Q_B (B)$. So, the dependence of source function on magnetic field gets factored out. The assumptions which we made allow us to find a solution which does not depend on initial magnetic fields. 

Of course, the main assumption here is that luminosity visibility function has a specific form. This assumption is widely used in literature \citep{2013IJMPD..2230021B}, and is observationally motivated. But one needs to keep in mind, that observationally motivated visibility function is effectively averaged over all angles and magnetic fields. More careful analysis requires the knowledge of the fraction of total energy losses which goes into radio emission. Unfortunately, the accurate model of radio emission is yet to be discovered. However, the fact that magnetic fields get factored out will remain true for any visibility function which has a form $V = V_{ P,\chi,\xi} V_B$, which allows for wide range of possible functions.

On the other hand, these considerations do not take into account the pulsar death line. The death line depends on pulsar parameters $P$, $\chi$ and $B$ in a way which does not allow to factor out magnetic field birth function. While the death line is not important for alignment model \citep{2014MNRAS.443.1891G}, it is very important for counter-alignment model \citep{BGI}. The reason for that is that it acts mostly on pulsars with large inclination angles. In MHD model such pulsars are young and energetic, and thus are not affected by the death line. In BGI model the situation is opposite. This is the reason why we can not consistently investigate DP interpulses in BGI model within the approach under consideration.
 
 \subsection{Number of interpulse pulsars}
 \label{sect:interpulse_fractions}
 
 We are now in a position to calculate the fraction of pulsars which have interpulse emission. Using solutions (\ref{MHD1})-(\ref{MHD2}) and (\ref{BGI1})--(\ref{BGI2}), visibility functions (\ref{eq:vis_DP}) and (\ref{eq:vis_SP}), as well as normalization (\ref{norm2}), we obtain number of single-pole and double-pole interpulses for a variety of models. The results are collected in Table \ref{table03}. 
 
  For MHD model we mostly use $Q_\chi = \sin \chi$, which gives angular distribution function in agreement with observations. For comparison, we also tried $Q_\chi = 2/\pi$ for $W_0 = 5.8^\circ P^{-1/2}$. One can see that because the solution (\ref{MHD1}) is divergent at small angles, the number of single-pole interpulses becomes unreasonably large. The number of double-pole interpulses is not so sensitive to $Q_\chi$ as is already seen from Figure \ref{fig:chi-dist-all}. We can thus conclude, that observations of pulsars at small inclination angles require `random' birth function $Q_\chi = \sin \chi$.
  
  For BGI model we use $Q_\chi = 2/\pi$, which also gives flat angular distribution function at small angles. Random birth function $Q_\chi = \sin \chi$ gives diminishing distribution function $N(\chi)$ and thus very small number of single-pole interpulses. 
 
 In addition to different evolution models and different inclination angle birth functions $Q_\chi$, we consider different window widths $W_0$. The value $W_0 = 2.45^\circ P^{-1/2}$ corresponds to the core component of radio emission \citep{1990ApJ...352..247R}, while the value $W_0 = 5.8^\circ P^{-1/2}$ describes the conal component \citep{1993ApJ...405..285R}. The latter value is not very well constrained due to lack of good statistics, so we use additional value $W_0 = 7.0^\circ P^{-1/2}$ to better constrain the number of interpulses (we note, that this value does not contradict observations of window widths). 
 
 For each window width we calculate upper and lower limits on single-pole interpulse fraction using the definitions from Section \ref{sect:vis_interpulses}. As could be easily shown, the number of such pulsars depends quadratically on the window width. For MHD model we get the best agreement with observations for $W_0 = 5.8^\circ P^{-1/2}$ with much worse agreement for other window widths. For BGI model, we obtain best agreement for $W_0 = 7.0^\circ P^{-1/2}$, and reasonable agreement for $W_0 = 5.8^\circ P^{-1/2}$.
 
 Thus, we can conclude that both models are able to describe the fraction of single-pole interpulses, and both of them require the use of visibility function for conal component (with BGI model requiring slightly larger window width).
 
 Our analysis allows us to estimate the number of double-pole interpulses only for MHD model. As a result, we get a fraction of such interpulses in a good agreement with observations. We obtain the best agreement for core component visibility function $W_0 = 2.45^\circ P^{-1/2}$. This fact is not surprising: the fit for core component of radio emission \citep{1990ApJ...352..247R,2012MNRAS.424.1762M} comes from the observations of DP interpulses (for which one can neglect $\sin^{-1}\chi$ factor in observed window width).

 \begin{table}
\caption{Prediction of the number of interpulse pulsars for MHD and BGI models. For single-pole interpulses we use criterion (\ref{eq:wj08}) to obtain lower limit, and relation (\ref{eq:circle}) for the upper limit. For double-pole interpulses there is no such uncertainty. For each model we try window widths $W_0$, corresponding to core and conal components of emission, as well as slightly larger value $W_0 = 7.0^\circ P^{-1/2}$. Unless mentioned in the third column, we use $Q_\chi = \sin\chi$ for MHD model and $Q_\chi = 2/\pi$ for BGI model. We can conclude that both evolution model are able to reproduce observations, although they require different birth functions.}
\centering
\begin{tabular}{lcl}

\multicolumn{3}{c}{Single-Pole interpulses ($0.03 ~{\rm s}\le P \le 0.5$ s)}\\
\hline
\hline
  Observations & $4 \div 14$& \\
MHD & $1 \div 3$ & \multirow{2}{*}{$W_0 = 2.45^\circ P^{-1/2}$}\\
BGI & $0.2 \div 0.6$\\
  \hline
MHD & $6 \div 18$ & \multirow{2}{*}{$W_0 = 5.8^\circ P^{-1/2}$}\\
BGI & $1 \div 4$\\
\hline
MHD & $9 \div 26$&\multirow{2}{*}{$W_0 = 7.0^\circ P^{-1/2}$}\\
BGI & $2 \div 5$\\ 
\hline
MHD & $17 \div 50$&\multirow{2}{*}{$W_0 = 10.0^\circ P^{-1/2}$}\\
BGI & $4 \div 12$\\ 
\hline
\hline
MHD & $19 \div 44$ & $Q_\chi = 2/\pi$\\
BGI & $0.08 \div 0.5$ & $Q_\chi = \sin \chi$\\ 
\hline
\vspace{0.2cm}\\
\multicolumn{3}{c}{Double-Pole interpulses ($0.03 ~{\rm s}\le P \le 0.5$ s)}\\
  \hline
  \hline
    Observations & $10 \div 23$ &\\
MHD & $15$&$W_0 = 2.45^\circ P^{-1/2}$\\

\hline
MHD & $36$&$W_0 = 5.8^\circ P^{-1/2}$\\
MHD & $44$&$W_0 = 7.0^\circ P^{-1/2}$\\
MHD & $28$&$Q_\chi = 2/\pi$\\
  \hline
\end{tabular}
\label{table03} 
\end{table}
 

 \subsection{Dependence on radio luminosity model}
 \label{sect:lum_funct}
 
 Even though we presented a solution of kinetic equation (\ref{eq:main}) only for luminosity visibility function $V_{\rm lum}^{\rm vis} = P^{-1}$, the results could be easily generalised for more sophisticated models. Indeed, any luminosity function of the form \mbox{$L \propto P^{\alpha_1} {\dot P}^{\alpha_2}$} can be expressed as 
 \begin{equation}
     L \propto P^{\kappa} f^{\rm lum}_{\chi}(\chi) f^{\rm lum}_{B}(B)\label{eq:Vlum_general}
 \end{equation}
with model-dependent $\kappa$, $f^{\rm lum}_{\chi}$, and $f^{\rm lum}_{B}$. For example, MHD model (\ref{MHD_P})--(\ref{MHD_chi}) has $\kappa = \alpha_1 - \alpha_2$, $f^{\rm lum}_{\chi} = (1+\sin^2\chi)^{\alpha_2}$, and $f^{\rm lum}_{B} = B^{2 \alpha_2}$, while BGI model (\ref{BGI_P})-(\ref{BGI_chi}) implies \mbox{$\kappa = \alpha_1 + \alpha_2/14$,} $f^{\rm lum}_{\chi} = \cos\chi^{2d\alpha_2}$, and $f^{\rm lum}_{B} = B^{10 \alpha_2/7}$.

In Table \ref{table04} we present the results for interpulse fraction for different luminosity models. We parametrize each model with power-law indices $\alpha_1$ and $\alpha_2$. The most widely used model (see Section \ref{sect:vis_f} for the discussion) is ($\alpha_1,~ \alpha_2)\sim (-1.5,~0.5)$ corresponds to the luminosity proportional to the potential drop over the polar cap. For comparison, we also include the model of constant fraction of radio luminosity in the total pulsar losses $L\propto I_r \Omega \dot \Omega \propto P^{-3}\dot P$. Finally, we use ($\alpha_1,~ \alpha_2)\sim (-0.8,~1/3)$ to elaborate analytical prediction (\ref{LBGI}). Formally, we can use this luminosity model only for BGI evolution theory. However, we include the results for MHD model for comparison as well.

We can conclude that the results depend significantly on the luminosity model. However, the conclusions of Section \ref{sect:interpulse_fractions} remain true for all luminosity models. Again, the uncertainties in both observations and theory prevent us from making exact evaluation of interpulse numbers. We are only able to make order of magnitude estimate. We see that for such estimates, we have an agreement for all models. However, with the growing number of observations, one will be required to have a good, physically-motivated luminosity model in order to obtain better agreement with observations.

\begin{table}
\caption{Prediction of the number of interpulse pulsars for 
 different radio luminosity models. We use $Q_\chi = \sin \chi$ for MHD model, and $Q_\chi = 2/\pi$ for BGI. One can see that the number of interpulse pulsars depends significantly on luminosity model. On the other hand, the conclusions of Section \ref{sect:interpulse_fractions} remain true for all models.} 
\centering
\begin{tabular}{lcc}
\multicolumn{3}{c}{Single-Pole interpulses, $W_0 = 5.8^\circ P^{-1/2}$}\\

\hline
\hline
  Observations & $4 \div 14$&$ (\alpha_1, \alpha_2)$\\

MHD & $1 \div 6$ &(-1.5,~0.5)\\
MHD & $0.5\div3 $ &(-3,~1)\\
MHD & $4 \div 13 $ &(-0.8,~1/3)\\
\hline
BGI &  $3 \div 9$   &(-1.5,~0.5)\\
BGI & $5 \div 15$ &(-3,~1)\\
BGI & $2 \div 7 $ & (-0.8,~ 1/3)\\
  \hline
\vspace{0.1cm}\\
\multicolumn{3}{c}{Double-Pole interpulses, $W_0 = 2.45^\circ P^{-1/2}$}\\
  \hline
  \hline
  MHD & $18$ & (-1.5,~0.5)\\
  MHD & $21$ & (-3,~1)\\
  MHD & $16$ & (-0.8, 1/3)\\
  \hline
  \vspace{0.1cm}\\
    \multicolumn{3}{c}{Double-Pole interpulses, $W_0 = 5.8^\circ P^{-1/2}$}\\
  \hline
  \hline
  Observations & $10 \div 23$&\\
  MHD & $44$ & (-1.5,~0.5)\\
  MHD & $51$ & (-3,~1)\\
  MHD & $40$ & (-0.8, 1/3)\\
  \hline
  
\end{tabular}
\label{table04} 
\end{table}
 
\section{Conclusions and discussion}
\label{sect:results}

Analysing now results collected in Table~\ref{table03}, one can conclude that both MHD model 
with the homogeneous birth function $Q_{\chi} = 2/\pi$ (\ref{MHD1}) as well as BGI model
for 'random' birth function $Q_{\chi} = \sin\chi $  (\ref{BGI2}) are in clear disagreement with
observations. The first model predicts too large number of SP interpulse pulsars while the 
second one predicts too low.  This result can be easily explained.

One can approximately evaluate the number of SP interpulses as
\begin{equation}
\label{eq:estimate}
    {\cal N}^{\rm SP} \sim {\cal N}_{\rm tot} W_0^2(P_{\rm med}) N_\chi^0/\langle N_\chi(\chi) \sin\chi \rangle_\chi,
\end{equation}
where $N_\chi^0$ is the characteristic value of angular distribution function near $\chi = 0$, which we take to be $N_\chi[W_0(P_{\rm med})]$, the denominator $\langle N_\chi(\chi) \sin\chi \rangle_\chi$ corresponds to the average value of angular distribution function of observed pulsars, which should be of order unity, unless most of the pulsars have small angles (for example, MHD model with uniform angular birth function), and $P_{\rm med}$ is the characteristic value of period, which we take to be 0.3~s.

For models which are in good agreement with observations (namely, MHD model with $Q_\chi = \sin\chi$ and BGI model with $Q_\chi = 2/\pi$), the estimate (\ref{eq:estimate}) gives
\begin{equation}
{\cal N}^{\rm SP}_{\rm MHD} \sim 0.05~{\cal N}_{\rm tot}, \qquad {\cal N}^{\rm SP}_{\rm BGI} \sim 0.01~{\cal N}_{\rm tot}.
\end{equation}
However, for MHD model with $Q_\chi = 2/\pi$, the same estimate gives ${\cal N}^{\rm SP}_{\rm MHD} \sim 0.1~{\cal N}_{\rm tot}$, which is too large. Similarly, BGI model with $Q_\chi = \sin \chi$ predicts ${\cal N}^{\rm SP}_{\rm BGI} \sim 0.002~{\cal N}_{\rm tot}$ which is too small.



Unfortunately, the precision of our considerations does not allow us to select the preferred model. On the other hand, our results put several constrains on the models. In particular, by selecting the model, one fixes the birth functions for the period range $0.03~{\rm s} \le P \le 0.5$~s:
\begin{enumerate}
\item MHD model requires random angular distribution function $Q_\chi = \sin \chi$. At the same time, this model requires the period birth function to be $Q_P = P^{-\kappa-1} \propto {\rm const}$.
\item BGI model requires uniform angular distribution function $Q_\chi = 2/\pi$. At the same time, this model requires the period birth function to be $Q_P = P^{-\kappa} \propto P$.
\end{enumerate}
Here we assume the simplest luminosity model with $\kappa = -1$. For both models the initial period distribution is rather broad. Our results in a good agreement with the results of \cite{2015ApJ...810..101F}, who computed initial spin periods of neutron stars which were spun up by internal gravity waves during core-collapse supernovae. On the other hand, stochastic spin up of the core will inevitably lead to random orientation of the angular momentum of the neutron start, and thus implies $Q_\chi = \sin\chi$, which is one of the requirements of MHD model.

In addition to predicting the number of interpulse pulsars, our method allows to determine the observed period distribution of interpulses. E.g., one can easily see from equations (\ref{eq:vis_SP}), (\ref{eq:vis_DP}) and (\ref{norm1}), that 
\begin{equation}
N^{\rm obs, SP} (P) \sim W_0^2(P) N^{\rm obs}(P) \propto P^{-1/2},
\end{equation}
and
\begin{equation}
N^{\rm obs, DP} (P) \sim W_0(P) N^{\rm obs}(P) \propto {\rm const}.
\end{equation}
Of course, the current number of observed interpulses is too small to compare their period distribution with our predictions. However, with the growing number of observed interpulses it will soon become possible.

Finally, as one can see from Table \ref{table03}, for single-pole interpulses the agreement with observations gets better with increasing window width. For double-pole interpulses the situation is opposite. This implies that the emission from low-inclination pulsars is dominated by the conal component, while the emission from high-inclination pulsars is mostly in core component.



To summarize, one can conclude that observational data are in agreement with both evolutionary
scenarios. Alignment MHD model predicting the reasonable number of both SP and DP interpulse pulsars.
As for counter-alignment BGI model, the analytical kinetic approach discussed above can 
give suitable number for SP interpulse pulsars only. To analyse DP interpulse pulsars, it is
necessary to include into consideration
\begin{enumerate}
\item
death line depending on magnetic field distribution of neutron stars (see Fig.~\ref{fig05}),
\item
the uncertainty in the antisymmetric current $i_{a}^{A}$ which describes the escaping rate for
orthogonal rotator through the death line,   
\item
inclination angle $\chi$ dependence of the radio luminosity $L^{\rm BGI}$ (\ref{LBGI}) resulting in 
the diminishing of the radio for orthogonal rotators.
\end{enumerate}
We are going to consider this case in the separate paper.



\section{Acknowledgments}

We thank A.V.~Biryukov, Ya.N.~Istomin, I.F.~Malov, E.B.~Nikitina, A.~Philippov, J.~Pons and S.B.~Popov  
for their interest and useful discussions. This work was partially supported by 
Russian Foundation for Basic Research (Grant no. 14-02-00831).


{\small
\bibliographystyle{mn2e}
\bibliography{mybib}

\begin{thebibliography}{}

\bibitem[\protect\citeauthoryear{{Arzamasskiy}, {Philippov} \&
  {Tchekhovskoy}}{{Arzamasskiy} et~al.}{2015}]{2015MNRAS.453.3540A}
{Arzamasskiy} L.,  {Philippov} A.,    {Tchekhovskoy} A.,  2015, \mnras, 453,
  3540

\bibitem[\protect\citeauthoryear{{Bagchi}}{{Bagchi}}{2013}]{2013IJMPD..2230021B}
{Bagchi} M.,  2013, International Journal of Modern Physics D, 22, 1330021

\bibitem[\protect\citeauthoryear{{Bai} \& {Spitkovsky}}{{Bai} \&
  {Spitkovsky}}{2010}]{2010ApJ...715.1282B}
{Bai} X.-N.,  {Spitkovsky} A.,  2010, \apj, 715, 1282

\bibitem[\protect\citeauthoryear{{Bates}, {Lorimer}, {Rane} \&
  {Swiggum}}{{Bates} et~al.}{2014}]{2014MNRAS.439.2893B}
{Bates} S.~D.,  {Lorimer} D.~R.,  {Rane} A.,    {Swiggum} J.,  2014, \mnras,
  439, 2893

\bibitem[\protect\citeauthoryear{{Beskin} \& {Eliseeva}}{{Beskin} \&
  {Eliseeva}}{2005}]{2005AstL...31..263B}
{Beskin} V.~S.,  {Eliseeva} S.~A.,  2005, Astronomy Letters, 31, 263

\bibitem[\protect\citeauthoryear{{Beskin}, {Gurevich} \& {Istomin}}{{Beskin}
  et~al.}{1984}]{1984Ap&SS.102..301B}
{Beskin} V.~S.,  {Gurevich} A.~V.,    {Istomin} I.~N.,  1984, \apss, 102, 301

\bibitem[\protect\citeauthoryear{{Beskin}, {Gurevich} \& {Istomin}}{{Beskin}
  et~al.}{1993}]{BGI}
{Beskin} V.~S.,  {Gurevich} A.~V.,    {Istomin} Y.~N.,  1993, {Physics of the
  pulsar magnetosphere}.
{Cambridge University Press}

\bibitem[\protect\citeauthoryear{{Beskin}, {Istomin} \& {Philippov}}{{Beskin}
  et~al.}{2013}]{2013PhyU...56..164B}
{Beskin} V.~S.,  {Istomin} Y.~N.,    {Philippov} A.~A.,  2013, Physics Uspekhi,
  56, 164

\bibitem[\protect\citeauthoryear{{Davis} \& {Goldstein}}{{Davis} \&
  {Goldstein}}{1970}]{1970ApJ...159L..81D}
{Davis} L.,  {Goldstein} M.,  1970, \apjl, 159

\bibitem[\protect\citeauthoryear{{Faucher-Gigu{\`e}re} \&
  {Kaspi}}{{Faucher-Gigu{\`e}re} \& {Kaspi}}{2006}]{2006ApJ...643..332F}
{Faucher-Gigu{\`e}re} C.-A.,  {Kaspi} V.~M.,  2006, \apj, 643, 332

\bibitem[\protect\citeauthoryear{{Fuller}, {Cantiello}, {Lecoanet} \&
  {Quataert}}{{Fuller} et~al.}{2015}]{2015ApJ...810..101F}
{Fuller} J.,  {Cantiello} M.,  {Lecoanet} D.,    {Quataert} E.,  2015, \apj,
  810, 101

\bibitem[\protect\citeauthoryear{{Goldreich}}{{Goldreich}}{1970}]{1970ApJ...160L..11G}
{Goldreich} P.,  1970, \apjl, 160, L11

\bibitem[\protect\citeauthoryear{{Good} \& {Ng}}{{Good} \&
  {Ng}}{1985}]{1985ApJ...299..706G}
{Good} M.~L.,  {Ng} K.~K.,  1985, \apj, 299, 706

\bibitem[\protect\citeauthoryear{{Gull{\'o}n}, {Miralles}, {Vigan{\`o}} \&
  {Pons}}{{Gull{\'o}n} et~al.}{2014}]{2014MNRAS.443.1891G}
{Gull{\'o}n} M.,  {Miralles} J.~A.,  {Vigan{\`o}} D.,    {Pons} J.~A.,  2014,
  \mnras, 443, 1891

\bibitem[\protect\citeauthoryear{{Lyne}, {Graham-Smith}, {Weltevrede},
  {Jordan}, {Stappers}, {Bassa} \& {Kramer}}{{Lyne}
  et~al.}{2013}]{2013Sci...342..598L}
{Lyne} A.,  {Graham-Smith} F.,  {Weltevrede} P.,  {Jordan} C.,  {Stappers} B.,
  {Bassa} C.,    {Kramer} M.,  2013, Science, 342, 598

\bibitem[\protect\citeauthoryear{{Lyne} \& {Graham-Smith}}{{Lyne} \&
  {Graham-Smith}}{1998}]{1998CAS....31.....L}
{Lyne} A.~G.,  {Graham-Smith} F.,  1998, Cambridge Astrophysics Series, 31

\bibitem[\protect\citeauthoryear{{Lyne}, {Manchester} \& {Taylor}}{{Lyne}
  et~al.}{1985}]{1985MNRAS.213..613L}
{Lyne} A.~G.,  {Manchester} R.~N.,    {Taylor} J.~H.,  1985, \mnras, 213, 613

\bibitem[\protect\citeauthoryear{{Maciesiak}, {Gil} \& {Melikidze}}{{Maciesiak}
  et~al.}{2012}]{2012MNRAS.424.1762M}
{Maciesiak} K.,  {Gil} J.,    {Melikidze} G.,  2012, \mnras, 424, 1762

\bibitem[\protect\citeauthoryear{{Maciesiak}, {Gil} \& {Ribeiro}}{{Maciesiak}
  et~al.}{2011}]{2011MNRAS.414.1314M}
{Maciesiak} K.,  {Gil} J.,    {Ribeiro} V.~A.~R.~M.,  2011, \mnras, 414, 1314

\bibitem[\protect\citeauthoryear{{Malov} \& {Nikitina}}{{Malov} \&
  {Nikitina}}{2013}]{2013ARep...57..833M}
{Malov} I.~F.,  {Nikitina} E.~B.,  2013, Astronomy Reports, 57, 833

\bibitem[\protect\citeauthoryear{{Manchester}, {Hobbs}, {Teoh} \&
  {Hobbs}}{{Manchester} et~al.}{2005}]{2005AJ....129.1993M}
{Manchester} R.~N.,  {Hobbs} G.~B.,  {Teoh} A.,    {Hobbs} M.,  2005, \aj, 129,
  1993

\bibitem[\protect\citeauthoryear{{Manchester} \& {Taylor}}{{Manchester} \&
  {Taylor}}{1977}]{1977puls.book.....M}
{Manchester} R.~N.,  {Taylor} J.~H.,  1977, {Pulsars}

\bibitem[\protect\citeauthoryear{{Mestel}, {Panagi} \& {Shibata}}{{Mestel}
  et~al.}{1999}]{MPS99}
{Mestel} L.,  {Panagi} P.,    {Shibata} S.,  1999, \mnras, 309, 388

\bibitem[\protect\citeauthoryear{{Philippov}, {Tchekhovskoy} \&
  {Li}}{{Philippov} et~al.}{2014}]{2014MNRAS.441.1879P}
{Philippov} A.,  {Tchekhovskoy} A.,    {Li} J.~G.,  2014, \mnras, 441, 1879

\bibitem[\protect\citeauthoryear{{Philippov}, {Spitkovsky} \&
  {Cerutti}}{{Philippov} et~al.}{2015}]{2015ApJ...801L..19P}
{Philippov} A.~A.,  {Spitkovsky} A.,    {Cerutti} B.,  2015, \apjl, 801, L19

\bibitem[\protect\citeauthoryear{{Popov} \& {Prokhorov}}{{Popov} \&
  {Prokhorov}}{2007}]{2007PhyU...50.1123P}
{Popov} S.~B.,  {Prokhorov} M.~E.,  2007, Physics Uspekhi, 50, 1123

\bibitem[\protect\citeauthoryear{{Popov} \& {Turolla}}{{Popov} \&
  {Turolla}}{2012}]{2012Ap&SS.341..457P}
{Popov} S.~B.,  {Turolla} R.,  2012, \apss, 341, 457

\bibitem[\protect\citeauthoryear{{Rankin}}{{Rankin}}{1990}]{1990ApJ...352..247R}
{Rankin} J.~M.,  1990, \apj, 352, 247

\bibitem[\protect\citeauthoryear{{Rankin}}{{Rankin}}{1993}]{1993ApJ...405..285R}
{Rankin} J.~M.,  1993, \apj, 405, 285

\bibitem[\protect\citeauthoryear{{Ruderman} \& {Sutherland}}{{Ruderman} \&
  {Sutherland}}{1975}]{1975ApJ...196...51R}
{Ruderman} M.~A.,  {Sutherland} P.~G.,  1975, \apj, 196, 51

\bibitem[\protect\citeauthoryear{{Smith}}{{Smith}}{1977}]{1977puls.book.....S}
{Smith} F.~G.,  1977, {Pulsars}.
Cambridge University Press

\bibitem[\protect\citeauthoryear{{Spitkovsky}}{{Spitkovsky}}{2006}]{2006ApJ...648L..51S}
{Spitkovsky} A.,  2006, \apjl, 648, L51

\bibitem[\protect\citeauthoryear{{Tauris} \& {Manchester}}{{Tauris} \&
  {Manchester}}{1998}]{1998MNRAS.298..625T}
{Tauris} T.~M.,  {Manchester} R.~N.,  1998, \mnras, 298, 625

\bibitem[\protect\citeauthoryear{{Tchekhovskoy}, {Philippov} \&
  {Spitkovsky}}{{Tchekhovskoy} et~al.}{2016}]{PTS16}
{Tchekhovskoy} A.,  {Philippov} A.,    {Spitkovsky} A.,  2016, \mnras, 457,
  3384

\bibitem[\protect\citeauthoryear{{Timokhin} \& {Arons}}{{Timokhin} \&
  {Arons}}{2013}]{2013MNRAS.429...20T}
{Timokhin} A.~N.,  {Arons} J.,  2013, \mnras, 429, 20

\bibitem[\protect\citeauthoryear{{Vivekanand} \& {Narayan}}{{Vivekanand} \&
  {Narayan}}{1981}]{1981JApA....2..315V}
{Vivekanand} M.,  {Narayan} R.,  1981, Journal of Astrophysics and Astronomy,
  2, 315

\bibitem[\protect\citeauthoryear{{Weltevrede} \& {Johnston}}{{Weltevrede} \&
  {Johnston}}{2008}]{2008MNRAS.387.1755W}
{Weltevrede} P.,  {Johnston} S.,  2008, \mnras, 387, 1755

\bibitem[\protect\citeauthoryear{{Young}, {Chan}, {Burman} \& {Blair}}{{Young}
  et~al.}{2010}]{2010MNRAS.402.1317Y}
{Young} M.~D.~T.,  {Chan} L.~S.,  {Burman} R.~R.,    {Blair} D.~G.,  2010,
  \mnras, 402, 1317

\bibitem[\protect\citeauthoryear{{Zanazzi} \& {Lai}}{{Zanazzi} \&
  {Lai}}{2015}]{2015MNRAS.451..695Z}
{Zanazzi} J.~J.,  {Lai} D.,  2015, \mnras, 451, 695

\end{thebibliography}
}
\bsp

\label{lastpage}

\end{document}